\documentclass[review,3p]{elsarticle}
\usepackage{graphicx}
\usepackage{bm}
\usepackage{pgfplots}
\usepackage{color}
\usepackage{algorithm}
\usepackage{amsmath,amssymb,amsthm,dsfont,mathrsfs}
\usepackage{multirow,epstopdf}

\journal{Journal of the Franklin Institute}

\def\diag{{\mathrm{diag}}}
\def\tr{{\mathrm{tr}}}
\def\A{{\mathbf{A}}}

\def\R{{\mathbf{R}}}
\def\I{{\mathbf{I}}}

\def\y{{\mathbf{y}}}

\def\v{{\mathbf{v}}}

\def\c{{\mathbf{c}}}
\def\Q{{\mathbf{Q}}}

\def\G{{\mathbf{G}}}

\def\e{{\mathbf{e}}}

\def\g{{\mathbf{g}}}
\def\G{{\mathbf{G}}}

\def\x{{\mathbf{x}}}
\def\Rc{{\mathbb{R}}}
\def\Cc{{\mathbb{C}}}
\def\Re{{\mathrm{Re}}}
\def\Im{{\mathrm{Im}}}

\begin{document}

\begin{frontmatter}
\title{Sparse Bayesian Learning Approach for Discrete Signal Reconstruction}

\cortext[mycorrespondingauthor]{Corresponding author}

\author[0,1]{Jisheng Dai}
\ead{jsdai@ujs.edu.cn}

\author[2]{An Liu}
\ead{anliu@zju.edu.cn}

\author[3]{Hing Cheung So}
\ead{hcso@ee.cityu.edu.hk}

\address[0]{College of Information Science and Technology, Donghua University, Shanghai 201620, China}
\address[1]{Department of Electronic Engineering, Jiangsu University, Zhenjiang 212013, China}
\address[2]{College of Information Science and Electronic Engineering, Zhejiang University, Hangzhou 310027, China}
\address[3]{Department of Electrical Engineering, City University of Hong Kong, Hong Kong, China}



\begin{abstract}
This study addresses the problem of discrete signal reconstruction from the perspective of sparse Bayesian learning
(SBL). Generally, it is intractable to perform the Bayesian inference with the ideal discretization prior under the SBL framework. To overcome this challenge, we introduce a novel discretization enforcing prior to exploit the knowledge of the discrete nature of the signal-of-interest.
By integrating the discretization enforcing prior into the SBL framework and applying the variational
Bayesian inference (VBI) methodology, we devise an alternating optimization algorithm to jointly characterize the finite-alphabet feature and reconstruct the
unknown signal. When the measurement matrix is i.i.d. Gaussian per component, we further embed the generalized  approximate message passing (GAMP) into the VBI-based method, so as to directly adopt the ideal prior and significantly reduce the computational burden. Simulation results
demonstrate substantial
performance improvement of the two proposed methods over existing schemes. Moreover, the GAMP-based
variant outperforms the VBI-based method with i.i.d. Gaussian measurement matrices but it fails to work for non i.i.d. Gaussian matrices.
\end{abstract}


\end{frontmatter}

\section{Introduction}

In the last decade, the problem of sparse signal recovery has attracted considerable attention in signal processing, and the associated compressed sensing (CS) technique \cite{donoho2006compressed,candes2008introduction} has been a paradigm for solving many important practical problems in a variety of fields, including radar \cite{malioutov2005sparse,liu2017off}, imaging processing \cite{lustig2008compressed,li2020nonconvex}, face recognition \cite{wright2009robust}, wireless communications \cite{kamel2022enhanced,rao2014distributed,zhou2022sparse}, and speech and audio processing \cite{plumbley2010sparse}.
Basically, CS aims to recover an unknown signal vector $\x$ that has only a few nonzero coefficients from an underdetermined  measurement.
There are two conventional classes of algorithmic approaches for CS, which are greedy pursuit  and convex relaxation \cite{souto2017efficient}. Greedy pursuit methods use a greedy strategy to determine the supports of $\x$ (e.g., orthogonal matching pursuit (OMP) algorithm \cite{pati1993orthogonal}); while convex relaxation methods try to  relax the nonconvex $l_0$-norm optimization problem into a convex one (e.g., the basis pursuit denoising (BPDN) algorithm \cite{chen2001atomic} or the $l_1$-norm minimization \cite{Donoho2008}). Although these methods work well for individual sparsity (the significant entries of $\x$ are assumed to be i.i.d.) and block sparsity (the significant entries of $\x$
cluster in blocks under a known specific sorting order), they cannot fully exploit additional  sparsity structures, e.g., burst sparsity and grouping sparsity, and they may suffer from a significant performance degradation due to any modeling mismatches.

Recently, sparse Bayesian learning (SBL) has become a very popular method for recovering sparse signals \cite{tipping2001sparse,ji2008bayesian,wipf2004sparse,xue2018new}, which adopts a hierarchical sparsity-enforcing prior to characterize the sparse signal from a Bayesian perspective. Compared with the $l_1$-norm minimization, the SBL-based framework can provide high flexibility to tackle the minimum $l_0$-norm problem \cite{ji2008bayesian,wipf2004sparse}, and it does not require the prior knowledge about the sparsity level, noise variance, and dictionary  mismatch, since it has an inherent learning capability through the Bayesian inference. Furthermore, it can provide a flexible way to deal with a variety of sparsity structures and/or  modeling mismatches.
For example,
burst-sparsity structure was exploited from the perspective of SBL to enhance the performance of sparse signal recovery in \cite{fang2015pattern,fang2016two,dai2019non}. Some structure-aware SBL methods or compressed sensing (CS) methods were also proposed to enhance the performance of sparse signal recovery \cite{liu2016exploiting,liu2018structure}.
The problem of joint signal recovery and common-sparsity grouping was first tackled in \cite{wang2016novel},
and then a more general sparsity model that has outliers deviated from the common-sparsity pattern was addressed in \cite{dai2019joint}. Dictionary refinement SBL-based methods for coping with modeling mismatches can be found in \cite{yang2013off,hu2012compressed,liu2013unified}.
All these studies have demonstrated that the SBL-based framework can significantly improve the recovery performance in  many practical scenarios, if more sophisticated sparsity structures and/or dictionary refinement techniques are exploited.

On the other hand, reconstructing discrete signals from incomplete linear measurements is also an important problem in signal processing.  Discrete signals taking values in a finite alphabet are very common in wireless communications, e.g., generalized spatial modulation \cite{liu2014denoising}, multiuser detection \cite{sasahara2017multiuser},  and cognitive spectrum sensing \cite{axell2012spectrum}, as well as discrete-valued image reconstruction \cite{duarte2008single,tuysuzoglu2015graph}. Since reconstructing an unknown discrete signal has a combinatorial nature, it will bring a NP-hard optimization problem whose computational time is exponential \cite{nagahara2015discrete}. If the discrete signal is sparse, we may apply a CS algorithm 
to obtain a sparse solution, and then project the solution onto the discrete set as in \cite{sparrer2014adapting}, but the performance of such separated operation is  not optimal.
Combining the sparsity and finite-alphabet property can improve the reconstruction performance \cite{tian2009detection,souto2017efficient,shim2014sparse}. However,  applying any existing CS algorithms to discrete signal reconstruction requires an additional assumption about  the finite alphabet, i.e., the finite alphabet should necessarily contain zero with a much higher probability than other entries of the finite alphabet. Such assumption about the finite alphabet may not always be valid in practice.

To reconstruct discrete signals with an arbitrarily finite alphabet, a new algorithm named
sum-of-absolute-values (SOAV) optimization has been proposed in \cite{nagahara2015discrete}.
The SOAV scheme belongs to the class of convex relaxation, which
relaxes the $l_0$-norm optimization problem with the use of $l_1$-norm.
Efficient algorithms based on proximal splitting \cite{combettes2011proximal} and  approximate message passing (AMP) \cite{donoho2009message} have been proposed for the SOAV optimization in \cite{sasahara2017multiuser,hayakawa2018discreteness}.
The asymptotic performance of discrete-valued vector reconstruction was analyzed  from linear measurement \cite{hayakawa2020asymptotic}.
Nevertheless, there are at least three limitations of the SOAV-based methods:
(i) it is designed for real-valued problems only;
(ii) suboptimal parameter selection and  $l_1$-norm convex relaxation might bring a performance loss;
and (iii) the  sparsity exploited in the $l_1$-norm minimization problem could be invalid for the finite alphabet with a large size.
To overcome these shortcomings, in this paper, we devise an SBL-based framework for general discrete signal reconstruction, as well as  a fast GAMP-based method if the measurement matrix is i.i.d. Gaussian per component.
Our contributions are summarized as follows:
\begin{itemize}
  \item {\bf Discretization Enforcing Prior}

  We introduce a novel discretization enforcing prior, which can exploit the knowledge of the discrete nature of the signal-of-interest (SOI). Compared with the ideal discretization prior, our discretization enforcing prior might bring a performance loss. However, since the ideal prior is composed of several Dirac delta functions, it is usually intractable to perform the Bayesian inference with the ideal prior. To overcome this challenge, we alternatively adopt a Gaussian distribution to approximate the Dirac delta function with a adjustable precision and assign a Gamma hyperprior for this precision.
  Such treatment is a commonly used trick for the SBL-based methods, which can provide a tractable Bayesian inference. To the best of our knowledge, this discretization enforcing prior has not been discussed for discrete signal reconstruction  in the literature.

  \item {\bf SBL-based Framework for Discrete Signal Reconstruction}

  We develop a general SBL-based framework for discrete signal reconstruction. The existing SBL-based methods were designed for the sparse signal recovery, and have not been applied for discrete signal reconstruction yet.
  To  jointly characterize the finite alphabet feature and reconstruct the unknown signal, we combine the discretization enforcing prior into the SBL-based framework, and then propose an algorithm with alternating updates based on
 the variational Bayesian inference (VBI) methodology \cite{tzikas2008variational} to perform the Bayesian inference.
 The proposed VBI-based method does not impose any restrictions on the measurement matrix.
 To the best of our knowledge, the SOAV optimization is the only method for discrete signal reconstruction with non i.i.d. Gaussian measurement matrices, but it has several shortcomings. Our VBI method can overcome
all the shortcomings of the SOAV optimization, and achieve the best performance for non i.i.d. Gaussian measurement matrices.
 When the measurement matrix is i.i.d. Gaussian distributed, we further embed the generalized approximate message passing (GAMP) \cite{rangan2010generalized}
into the VBI-based method  to propose a fast GAMP variant. 
It is worth mentioning that the application of VBI or GAMP aims to perform the Bayesian inference with the new discretization enforcing prior. Neither application is straightforward, because most of the
updating rules for the Bayesian inference are needed to be re-derived in the presence of the new discretization enforcing prior.


%

\end{itemize}


The rest of the paper is organized as follows. In Section II, we present the signal model and review the state-of-the-art SOAV optimization for discrete signal reconstruction.
In Section III, we devise the VBI-based method for discrete signal reconstruction. In Section IV, we
further develop the fast GAMP-based method. Numerical experiments and
discussions  follow in Sections V and VI, respectively.

$Notations:$ $\mathbb{C}$ denotes complex number, $\mathbb{R}$ denotes real number, $\|\cdot\|_p$ denotes $p$-norm, $(\cdot)^T$ denotes transpose, $(\cdot)^H$ denotes Hermitian transpose,
$\I_N$ denotes $N\times N$ identity matrix, $\bm{1}_N$ denotes $N\times 1$ vector with all entries being $1$,
$\propto$ denotes equality up to a multiplicative constant or an additive constant,
$\mathcal{CN}(\cdot  | \bm\mu, \bm\Sigma)$ denotes complex Gaussian distribution with mean $\bm\mu$ and covariance  $\bm\Sigma$, $\tr(\cdot)$ denotes trace operator, $\diag(\cdot)$  denotes diagonal operator,
$[\x=\y]$ indicates whether $\x$ is equal to $\y$ or not (which returns $1$ if $\x=\y$; otherwise $0$ is returned),
$\Re(\cdot)$ denotes real part, and $\Im(\cdot)$ denotes imaginary part.

\section{Data Model and Existing solutions}
In this section, we first present the data model for discrete signal reconstruction, and then review the SOAV optimization approach and its shortcomings.


\subsection{Data Model}
Consider the problem of recovering a complex-valued discrete vector $\x=[x_1,x_2,\ldots, x_N]^T\in\Cc^{N\times 1}$
from an underdetermined  measurement vector $\y=[y_1,y_2,\ldots,y_M]^T\in \Cc^{M\times 1}$:
\begin{align}\label{eq-csm}
\y=\A\x + \v,
\end{align}
where $\A\in \Cc^{M\times N}$  is a  sampling matrix with $M < N$, and $\v =[v_1,v_2,\ldots, v_M]^T\in \Cc^{M\times 1}$ stands for an additive complex i.i.d. Gaussian noise vector with zero-mean and variance $\sigma^2$ for each entry.
Assume that the elements of $\x$ are i.i.d. discrete variables  from  a given finite alphabet $\mathcal{F}=\{f_l\}_{l=1}^L$ with a prior distribution:
\begin{align}\label{eq-fa}
P(x_n=f_l)=\rho_l, ~~n=1,2,\ldots, N, ~l=1,2,\ldots,L,
\end{align}
where $f_l\in \Cc$, $\rho_l\ge 0$ and $\sum_{l=1}^L \rho_l=1$. Obviously,  the ideal discretization prior (\ref{eq-fa}) can be rewritten as
\begin{align}\label{eqideal0}
p(x_n)=\sum_{l=1}^L \rho_l \delta(x_n-f_l),
\end{align}
where $\delta(\cdot)$ stands for the Dirac delta function. The
maximum $a$ $posterior$ (MAP) estimate of $\x$ is then:
\begin{align}\label{mapx}
\x^\star =  \arg \max_{\x} p(\x|\y)   =\arg \max_{\x} \left( p(\y|\x) \cdot \prod_{n=1}^N p(x_n) \right).
\end{align}
Since each $p(x_n)$ contains $L$ Dirac delta functions,
computing $\x^\star$ requires a combinatorial search.
Therefore,  (\ref{mapx}) is a NP-hard optimization
problem whose computational complexity  is exponential.


\subsection{SOAV Optimization}

The SOAV optimization is the state-of-the-art method for real-valued discrete signal reconstruction. In the following, we first review the SOAV optimization for  the real-valued discrete signal reconstruction, and then discuss how to extend it to handle complex-valued problems.

Assume that all the terms in (\ref{eq-csm}) are real-valued (i.e., $\y\in \Rc^{M\times 1}$, $\A\in \Rc^{M\times N}$, $\x\in \Rc^{N\times 1}$ and $\v\in \Rc^{M\times 1}$) and $f_l\in \Rc, \forall l$.
The SOAV optimization notices that the  vector $(\x-f_{l}\cdot\bm{1}_N)$ has approximately $\rho_{l}N$ zero elements. Taking advantage of the CS paradigm, the real-valued discrete vector $\x$ can be obtained by \cite{hayakawa2018discreteness}
\begin{align}\label{SOAV1}
\min_{\x} \sum_{l=1}^{L} \eta_{l} \|\x-f_l\cdot\bm{1}_N\|_1   +  \frac{\lambda}{2}\|\y- \A\x\|_2^2,
\end{align}
where $\lambda >0 $ is a regularization parameter which maintains a proper balance between empirical loss and regularization level,
and the coefficient $\eta_{l}\ge 0$ is fixed as $\eta_l=\rho_{l}$ in \cite{nagahara2015discrete} and $\eta_{l}=1$ in \cite{aissa2015sparsity}.
Note that the solution to (\ref{SOAV1}) is exactly equal to the MAP estimate with the prior distribution
$p(\x)\propto\exp(-\sum_{l=1}^{L} \eta_{l}\|\x-f_{l}\cdot\bm{1}_N\|_1)$.
Proximal-splitting-based algorithm \cite{sasahara2017multiuser} and AMP-based algorithm \cite{hayakawa2018discreteness} have been proposed for solving the $l_1$-norm minimization problem (\ref{SOAV1}).
It has been demonstrated in \cite{hayakawa2018discreteness} that the performance can be improved if $\eta_{l}$s are also considered as parameters to be optimized.

We may extend the   SOAV optimization methods to handle the complex-valued problem as in \cite{nagahara2015discrete,sasahara2017multiuser,hayakawa2018discreteness}.
Specifically, the complex-valued signal model (\ref{eq-csm}) is transformed  into a real-valued model as
\begin{align}
\begin{bmatrix}
\Re(\y)\\
\Im(\y)
\end{bmatrix}
=
\begin{bmatrix}
\Re(\A) ~ -\Im(\A)\\
\Im(\A) ~~~~ \Re(\A)
\end{bmatrix}
\begin{bmatrix}
\Re(\x)\\
\Im(\x)
\end{bmatrix}
+
\begin{bmatrix}
\Re(\v)\\
\Im(\v)
\end{bmatrix},\label{eq-rel}
\end{align}
and then  SOAV optimization can be applied to
such real-valued model. Actually, a natural extension is to directly use  (\ref{SOAV1}) by replacing the real vectors/matrices with complex ones. However, the complex-valued form of  (\ref{SOAV1}) prevents the SOAV optimization methods from adopting the proximal splitting algorithm proposed in \cite{sasahara2017multiuser,hayakawa2018discreteness}.
On the other hand, a complex AMP algorithm was proposed in \cite{jeon2015optimality} for the complex discrete-valued vector reconstruction, but it only works for i.i.d. Gaussian measurement matrices.

\subsection{Shortcomings for SOAV Optimization}

The main shortcomings of the SOAV optimization are:
\begin{itemize}
  \item Although the real-valued SOAV optimization methods can be used for the complex-valued discrete signal reconstruction by transforming the complex-valued model (\ref{eq-csm}) into the equivalent real-valued model (\ref{eq-rel}), but it cannot handle dependent real and imaginary parts \cite{hayakawa2018discreteness}. What is worst, the size of the finite alphabet  may become twice in the worst case (i.e., $\Re(\mathcal{F}) \cup \Im(\mathcal{F})$), which may cause a substantial performance degradation due to the possible nearby elements in the finite alphabet.

  \item The standard $l_1$-norm minimization formulation  (\ref{SOAV1})  used for the SOAV optimization will bring a performance loss, because (i) $l_1$-norm is a simple approximation of $l_0$-norm, which has a worse approximation performance than SBL \cite{wipf2004sparse}; and (ii) the regularization term $\lambda$ is regarded as a genuine nuisance parameter, and we usually select its suboptimal value only.

  \item The SOAV optimization exploits the sparsity from the fact that the vector $(\x-f_{l}\cdot\bm{1}_N)$ has approximately $\rho_{l}N$ zero elements. However, for the  finite alphabet $\mathcal{F}$ with a large size, the value of $\rho_{l}$ could be quite small, and thus the sparsity of the  vector $(\x-f_{l}\cdot\bm{1}_N)$ is hard to be guaranteed in this case.

\end{itemize}

To address the above issues,
we will  directly take knowledge of the discrete nature of the signal into account inside the SBL framework, and
devise an VBI-based approach for general discrete signal reconstruction, as well as a fast GAMP  variant  when all entries in the measurement matrix $\A$ are i.i.d. Gaussian distributed.

\section{VBI for Discrete Signal Reconstruction}

\subsection{New Discretization Enforcing Prior}

As will be shown later, it is usually intractable to perform the Bayesian inference with  the ideal discretization prior (\ref{eqideal0}). Hence, in this subsection, we design a novel discretization enforcing prior, which can exploit the knowledge of the discrete nature of SOI under the SBL framework.

\noindent\textbf{Definition~1.} \emph{Discretization Enforcing Prior}: Let $\gamma_n$ be the precision of $x_n$ and
$\g_n=[g_{n,1}, g_{n,2},\ldots, g_{n,L}]^T$ be an assignment vector that takes values from $\e_1, \e_2,\ldots, \e_L$,
where $\e_l$ stands for an $L\times 1$ zero vector except for the $l$-th element being 1, then we model the distribution of $x_n$ conditional on $\g_n$ and $\gamma_n$ as
\begin{align}\label{eq-prix}
p(x_n|\g_n,\gamma_n)= \prod_{l=1}^L  \left\{\mathcal{CN}(x_n |  f_l  , \gamma_n^{-1} )\right\}^{g_{n,l}},
\end{align}
where $\gamma_n$  is further modeled as a Gamma hyperprior
\begin{align}\label{eq-prigam}
p(\gamma_n)=& \Gamma(\gamma_n|a,b)
\end{align}
with $a$ and $b$ being some small constants (e.g., $a=b=10^{-6}$).

Due to the introduction of the assignment vector  $\g_n$ and different discrete means $f_l$s into the Gaussian distribution,
the two-stage hierarchical prior (\ref{eq-prix}) and (\ref{eq-prigam}) can  exploit the knowledge of the discrete nature of $x_n$ as follows.
Without loss of generality, let $\g_n=\e_l$ and then we have
\begin{align}\label{Let1}
 p(x_n|\g_n=\e_l, \gamma_n) = \mathcal{CN}(\underbrace{x_n- f_l}_{\triangleq \tilde{x}_{n,l}} |  0  , \gamma_n^{-1} ),
\end{align}
which follows the definition (\ref{eq-prix}) directly. With (\ref{Let1}), we obtain
\begin{align}
p(x_n|\g_n=\e_l)=&\int_0^\infty p(x_n|\g_n=\e_l,\gamma_n) p(\gamma_n) d\gamma_n\notag\\
=& \int_0^\infty \mathcal{CN}(\tilde{x}_{n,l} |  0  , \gamma_n^{-1} ) \Gamma(\gamma_n|a,b) d\gamma_n\notag\\
\propto & \left(b +  |\tilde{x}_{n,l}|^2 \right)^{-(a+\frac{1}{2})}.\label{Lemt3}
\end{align}
Here, we use  (42) in \cite{wipf2004sparse} to derive (\ref{Lemt3}).
Clearly, $p(x_n|\g_n=\e_l)$ is proportional to a Student-t distribution. Since $b$ is allowed to be very small, $p(x_n|\g_n=\e_l)$  is recognized as encouraging sparsity of $\tilde{x}_{n,l}$ \cite{wipf2004sparse}, which, in return, enforces  $x_n\rightarrow f_l$.
It is worth noting that the  discrete signal $\x$ is non-sparse itself and the discretization enforcing prior is used to exploit the discrete nature of $\x$ rather than its sparsity.


If the distribution of the finite alphabet $\mathcal{F}$ is available, the prior distribution of $\g_n$ can be formulated as a categorical distribution:
\begin{align}\label{disg}
p(\g_n)= \prod_{l=1}^L \rho_l^{[\g_n=\e_l]},
\end{align}
or, equivalently\footnote{This equivalence follows from the fact that
only one element of  the assignment vector $\g_n$ is activated on a single trial.
For example, if $\g_n=\e_1$, both (\ref{disg}) and (\ref{disg2}) give the same value
$(\rho_1)^1  (\rho_2)^0 (\rho_3)^0\ldots (\rho_L)^0$.}
\begin{align}\label{disg2}
p(\g_n)= \prod_{l=1}^L  \rho_l^{g_{n,l}},
\end{align}
where $\rho_{l}$ stands for the probability  of the $l$-th element in the finite alphabet $\mathcal{F}$.
Otherwise, it may be formulated as a non-informative distribution:
\begin{align}\label{eq-progg}
p(\g_n)= \prod_{l=1}^L \left(\frac{1}{L}\right)^{g_{n,l}}.
\end{align}
Note that the categorical distribution (\ref{disg}) is a special case of the multinomial distribution, which gives the
probabilities of potential outcomes of a single drawing only.
When the entries of both $\x$ and $\bm\gamma=[\gamma_1,\gamma_2,\ldots, \gamma_N]^T$ are i.i.d., we have
\begin{align}
p(\x|\G,\bm\gamma)=& \prod_{n=1}^N  \prod_{l=1}^L  \left\{\mathcal{CN}(x_n |  f_l  , \gamma_n^{-1} )\right\}^{g_{n,l}},\label{eq-pA1}\\
p(\bm\gamma)=& \prod_{n=1}^N\Gamma(\gamma_n|a,b),\label{eq-pA2}
\end{align}
where $\G=\{\g_n\}_{n=1}^N$.

\noindent\textbf{Remark~1.} Recall that the original SBL \cite{tipping2001sparse} adopts the  well-known Gaussian mixture as the prior, i.e.,
\begin{align}
p(x_n|\gamma_n)
=  \mathcal{CN}(x_n | 0 , \gamma_n^{-1} )\label{eq-pB1}
\end{align}
and
\begin{align}\label{eq-prigam1}
p(\gamma_n)=& \Gamma(\gamma_n|a,b).
\end{align}
Compared with the above prior, our devised two-stage hierarchical prior in (\ref{eq-prix}) and  (\ref{eq-prigam}) can be seen as its extension, and includes it  as a special case if $L=1$ and $f_1=0$.
The newly introduced assignment variables $\g_n$s will be automatically learned by the Bayesian inference,  allowing different elements of the finite alphabet to adaptively focus on different parts of the discrete SOI. In this case, the proposed method based on the discretization enforcing prior will significantly improve the discrete signal reconstruction performance, in contrary to placing fixed weight on each element of the finite alphabet in (\ref{SOAV1}).

%

\noindent\textbf{Remark~2.} The ideal value of each $\gamma_n$ should be infinite, since every $x_n$ exactly takes value in the finite alphabet $\mathcal{F}$ [see (\ref{eq-fa})]. In this case, the distribution $p(x_n|\g_n,\gamma_n)$ reduces to
\begin{align}\label{fa2}
p(x_n|\g_n)= \prod_{l=1}^L  \left\{ \delta(x_n-f_l)     \right\}^{g_{n,l}},
\end{align}
which is equal to the  ideal discretization prior (\ref{eqideal0}), becasue
\begin{align}
\sum_{\g_n\in \{\e_l  \}_{l=1}^L  } p(x_n|\g_n) p(\g_n)= \sum_{l=1}^L \rho_l \delta(x_n-f_l).
\end{align}
Unfortunately,
as will be shown later, it is intractable to perform the Bayesian inference with (\ref{fa2}).
Hence, we alternatively consider $\gamma_n$ as a variable and assign a Gamma hyperprior for it as in Definition~1. Such treatment is a commonly used trick for the SBL-based methods, because a Gamma distribution is a conjugate prior of a Gaussian distribution, which can provide a tractable Bayesian inference.
Empirical evidence shows that $\gamma_n$s will be automatically set to some large values through
the Bayesian inference.

\noindent\textbf{Remark~3.} In the next section, we will show that it is possible to adopt the ideal prior (\ref{fa2}) directly, if the marginal posterior $p(x_n|\y),\forall n$, can be approximately calculated.
However, this approximation requires the assumption that the elements of the measurement matrix $\A$ are i.i.d. Gaussian distributed. Without such assumption, the approximation method in Section IV might give a very bad performance; while the VBI-based method with the new discretization enforcing prior does not impose any assumption about $\A$.

\subsection{Proposed VBI-based Method}
Utilizing the new discretization enforcing prior presented in Definition~1, we will develop a general VBI-based method for discrete signal reconstruction in this subsection.
Under the assumption of the additive complex i.i.d. Gaussian noises, we have
\begin{align}\label{eq-noipro}
p(\y|\x,\alpha)= \mathcal{CN}(\y | \A\x, \alpha^{-1}\I  ),
\end{align}
where $\alpha=\sigma^{-2}$ stands for the noise precision, which can be similarly  modeled as in (\ref{eq-prigam})
\begin{align}\label{eq-alphapro}
p(\alpha)= \Gamma(\alpha|a,b).
\end{align}
Let $\bm\Omega \triangleq \{\alpha, \x, \bm\gamma, \G  \}$ be the set of hidden variables to  be estimated, and then the joint distribution $p(\y, \bm\Omega)$ can be expressed as
\begin{align}\label{eq-jprob}
p(\y, \bm\Omega)= p(\y|\x,\alpha) p(\x|\G,\bm\gamma)  p(\alpha) p(\bm\gamma) p(\G).
\end{align}
If we could calculate the MAP estimate of $\bm\Omega$ from $p(\bm\Omega|\y) =  p(\y,\bm\Omega)/ p(\y)$, i.e.,
\begin{align}\label{eq-posterior}
\bm\Omega^\star =\max_{\bm\Omega} p(\bm\Omega|\y) = \max_{\bm\Omega} p(\y,\bm\Omega),
\end{align}
the finite-alphabet feature and unknown signal will be jointly obtained. To determine the final discrete signal, we may either project the MAP estimate of $\x$ onto the discrete set $\mathcal{F}$, or find the maximum element of the MAP estimate of $\g_n$. Both operations give very a similar estimation performance, but we prefer the second one because it is much simpler than the first.
Nevertheless,
it is very challenging to solve the problem (\ref{eq-posterior}) directly.
The VBI methodology \cite{tzikas2008variational}  is the state-of-the-art approach handling
intractable MAP estimate problems, which aims to find a simple approximate posterior instead of the true posterior.
Besides it, only numerical methods (e.g., Markov chain
Monte Carlo (MCMC) method and Gibbs sampling) are available in the literature.
Following the main procedures adopted in our previous work \cite{dai2019non}, we propose an VBI-based method to jointly exploit the finite-alphabet feature and reconstruct the unknown signal.


The basic idea of the VBI methodology is to find an approximate posterior $q(\bm\Omega)$ instead of $ p(\bm\Omega|\y)$. Here, we adopt the mean field approximation:
\begin{align}\label{eq-vbiform}
q(\bm\Omega)= q(\alpha) q(\x) q(\bm\gamma) q(\G)
\end{align}
which can make the approximate posterior analytically tractable \cite{tzikas2008variational}, but it is not the only way to perform the factorization.
The \lq\lq best" solution under the factorized constraint in (\ref{eq-vbiform}) should have the minimum Kullback-Leibler (KL) divergence between  $q(\bm\Omega)$  and $ p(\bm\Omega|\y)$, i.e.,
\begin{align}\label{eq-KLproblem}
q^\star(\bm\Omega) =  \min_{q(\bm\Omega)} D_{\mathrm{KL}}(q(\bm\Omega )|| p(\bm\Omega| \y)  ),
\end{align}
where $ D_{\mathrm{KL}} (q(x)||p(x)) \triangleq\int  q(x )  \ln \frac{ q(x)}{p(x)} dx$.
As shown in \cite{tzikas2008variational,dai2019non}, the optimal solution to  (\ref{eq-KLproblem}) should satisfy the following equality
\begin{align}\label{eq-slu}
\ln q^\star (\Omega_k)\propto \left< \ln p(\y,\bm\Omega  ) \right>_{\prod_{j\ne k} q^{\star}(\Omega_j)} ,~k=1,2,3,4,
\end{align}
where $\Omega_k$ stands for the $k$-th element in $\bm\Omega$. Note that each solution $q^\star (\Omega_k)$ given in (\ref{eq-slu}) is dependent on others ($q^\star (\Omega_j), j\ne k$). Therefore,
it is intractable to find the optimal closed-form solution.
Following the alternating optimization algorithm proposed in \cite{dai2019non}, a stationary solution can be found instead by iteratively  updating $q(\alpha)$, $q(\x)$, $q(\bm\gamma)$ and $q(\G)$ as:
\begin{align}
\ln q^{(i+1)}(\alpha)&\propto  \left< \ln p (\y,\bm\Omega  ) \right>_{q^{(i)}(\x)q^{(i)}(\bm\gamma)q^{(i)}(\G)},\label{eqM1}\\
\ln q^{(i+1)}(\x)&\propto  \left< \ln p(\y,\bm\Omega  ) \right>_{q^{(i+1)}(\alpha)q^{(i)}(\bm\gamma)q^{(i)}(\G)},\label{eqM2}\\
\ln q^{(i+1)}(\bm\gamma)&\propto  \left< \ln p(\y,\bm\Omega  ) \right>_{q^{(i+1)}(\alpha)q^{(i+1)}(\x)q^{(i)}(\G)},\label{eqM3}\\
\ln q^{(i+1)}(\G)&\propto  \left< \ln p(\y,\bm\Omega  ) \right>_{q^{(i+1)}(\alpha)q^{(i+1)}(\x)q^{(i+1)}(\bm\gamma)},\label{eqM4}
\end{align}
where 
$(\cdot)^{(i)}$ denotes the $i$-th iteration. In the following, we will address the updates (\ref{eqM1})--(\ref{eqM4}) in detail, and discuss the convergence of the proposed algorithm.

\subsection{Detailed Updates for (\ref{eqM1})--(\ref{eqM4})}
In this subsection, we focus on dealing with the updates for $q(\alpha)$, $q(\x)$, $q(\bm\gamma)$ and $q(\G) $.  Note that the update for $q(\alpha)$ coincides with the one in \cite{dai2019non} due to using the same Gaussian noise model, but the updates for $q(\x)$, $q(\bm\gamma)$ and $q(\G) $ are different because of adopting the new discretization enforcing prior (\ref{eq-prix}).

\subsubsection{Update of $q(\alpha)$} According to (\ref{eqM1}) and (\ref{eq-jprob}),
\begin{align}\label{eq-ut1}
\ln q^{(i+1)}(\alpha) \propto  \left<  \ln p(\y|\x, \alpha) p(\alpha) \right>_{q^{(i)}(\x)}.
\end{align}
Substituting (\ref{eq-noipro}) and (\ref{eq-alphapro}) into (\ref{eq-ut1})  yields
\begin{align}
&\ln q^{(i+1)}(\alpha)\notag\\
\propto& (a + M -1)\ln \alpha -  \alpha\cdot\left( b+   \left<\|\y - \A \x \|_2^2 \right>_{q^{(i)}(\x)} \right)\notag \\
 \propto& (a + M -1)\ln \alpha -  \alpha\cdot\left( b+  \| \y - \A \bm\mu^{(i)}    \|_2^2     + \tr(\A \bm\Sigma^{(i)}\A^H) \right), \label{eq-upal01}
\end{align}
where $\bm\mu^{(i)} \triangleq \left<  {\x} \right>_{q^{(i)}(\x)} $ and $\bm\Sigma^{(i)} \triangleq  \left<  ({\x} - \bm\mu^{(i)} )({\x} - \bm\mu^{(i)} )^H\right>_{q^{(i)}(\x)}$. Hence, $q^{(i+1)}(\alpha)$ obeys a Gamma distribution
\begin{align}\label{eq-alg1}
q^{(i+1)}(\alpha)=& \Gamma(\alpha| a + M , b_\alpha^{(i+1)}  ),
\end{align}
where $ b_\alpha^{(i+1)} = \| \y - \A \bm\mu^{(i)}    \|_2^2     + \tr(\A \bm\Sigma^{(i)}\A^H) $.

\subsubsection{Update of $q(\x)$} The update (\ref{eqM2}) leads to
\begin{align}
&\ln q^{(i+1)}(\x)\notag\\
& \propto  \left<   \ln p( \y | \x, \alpha ) p(\x|\G,\bm\gamma)        \right>_{q^{(i+1)}(\alpha) q^{(i)}(\bm\gamma) q^{(i)}(\G) }. \label{eq-ut2}
\end{align}
Substituting (\ref{eq-noipro}) and (\ref{eq-pA1}) into (\ref{eq-ut2}), we have
\begin{align}
&\ln q^{(i+1)}(\x)\notag\\
\propto&  -\hat{\alpha}^{(i+1)}\| \y- \A\x \|_2^2     -\sum_{n=1}^N \sum_{l=1}^L  \phi_{n,l}^{(i)} \hat\gamma_n^{(i)} |x_n-f_l|^2          \\
\propto&-\hat{\alpha}^{(i+1)}\| \y- \A\x \|_2^2
-\sum_{l=1}^L  (\x- f_l\cdot \mathbf{1}  )^H \Q_l^{(i)} ( \x- f_l\cdot \mathbf{1} ),  \label{eq-ut2-1}
\end{align}
where $\hat{\alpha}^{(i+1)}=\left< \alpha\right>_{q^{(i+1)}(\alpha)}$,
$\phi_{n,l}^{(i)}\triangleq q^{(i)}(\g_{n} = \e_l )$,  $\hat\gamma_n^{(i)} = \left< \gamma_n\right>_{q^{(i)}(\gamma_n )}$, $\Q_l^{(i)}=\diag\{   \phi_{1,l}^{(i)} \hat\gamma_1^{(i)} ,   \phi_{2,l}^{(i)} \hat\gamma_2^{(i)},\ldots,   \phi_{N,l}^{(i)} \hat\gamma_N^{(i)}    \} $.  According to (\ref{eq-ut2-1}), $q^{(i+1)}(\x)$ should obey a Gaussian distribution:
\begin{align}\label{eq-alg2}
q^{(i+1)}(\x)=  \mathcal{CN}(\x| \bm\mu^{(i+1)} ,  \bm\Sigma^{(i+1)}   ),
\end{align}
where
\begin{align}
 \bm\mu^{(i+1)}=&  \bm\Sigma^{(i+1)}\left(  \hat{\alpha}^{(i+1)} \A^H\y +  \sum_{l=1}^L f_l\cdot \mathbf{1}^H \Q_l^{(i)}  \right),\\
  \bm\Sigma^{(i+1)} =&  \left( \hat{\alpha}^{(i+1)} \A^H\A + \sum_{l=1}^L \Q_l^{(i)} \right)^{-1} .\label{eqSinv}
\end{align}

It is straightforward to extend the above updating rule to the case that $\x$ is  factorized independently in each element.
The element-independent factorization method is called the space alternating variational estimation (SAVE)
method in \cite{thomas2018save}. Note that SAVE can provide an efficient Bayesian inference by avoiding the matrix inverse in (\ref{eqSinv}), but it suffers from a performance loss (as will be shown in the simulations later), because it adopts more approximate operations.


\subsubsection{Update of $q(\bm\gamma)$}  According to (\ref{eqM3}) and (\ref{eq-jprob}),  we have
\begin{align}\label{eq-ut3}
\ln q^{(i+1)}(\bm\gamma) \propto  \left<  p(\x|\G,\bm\gamma)  p(\bm\gamma)\right>_{q^{(i+1)}(\x) q^{(i)}(\G)  }.
\end{align}
Substituting (\ref{eq-pA1}) and (\ref{eq-pA2}) into (\ref{eq-ut3}) yields
\begin{align}
&\ln q^{(i+1)}(\bm\gamma)\notag\\
  \propto&
 \sum_{n=1}^N \left( a + \sum_{l=1}^L \phi_{n,l}^{(i)} -1 \right)\ln\gamma_n -\sum_{n=1}^N \gamma_n \left( b+ \sum_{l=1}^L  \phi_{n,l}^{(i)}\left< \left|x_n- f_l\right|^2  \right>_{q^{(i+1)}(x_n)}  \right)\\
 \propto&
 \sum_{n=1}^N \left( (a + 1) -1 \right)\ln\gamma_n  -\sum_{n=1}^N \gamma_n \left( b+ \sum_{l=1}^L  \phi_{n,l}^{(i)} \chi_{n,l}^{(i+1)}    \right),  \label{eq-ut3-1}
\end{align}
where $\chi_{n,l}^{(i+1)} = \left< \left|x_n-f_l\right|^2  \right>_{q^{(i+1)}(x_n)} $.
Since the terms related to each $\gamma_n$ are separable  in (\ref{eq-ut3-1}), $q^{(i+1)}(\gamma_{n})$ should obey a Gamma distribution:
\begin{align}\label{eq-alg3}
q^{(i+1)}(\gamma_{n})= \Gamma\left(\gamma_{n} |a+1, b_{n}^{(i+1)}\right),
\end{align}
where $b_{n}^{(i+1)} =  b+ \sum_{l=1}^L  \phi_{n,l}^{(i)}\chi_{n,l}^{(i+1)}$.

\subsubsection{Update of $q(\G)$} The update (\ref{eqM4}) leads to
\begin{align}
\ln q^{(i+1)}(\G)\propto   \left<\ln p(\x|\G,\bm\gamma) p(\G) \right>_{q^{(i+1)}(\x)q^{(i+1)}(\bm\gamma)}.\label{eq-ut4}
\end{align}
Substituting (\ref{eq-pA1}) and (\ref{disg2}) into (\ref{eq-ut4}), we obtain
\begin{align}
&\ln q^{(i+1)}(\G)=  -\sum_{n=1}^N \sum_{l=1}^L g_{n,l} \hat\gamma_n^{(i+1)} \left( \left|\mu_n^{(i+1)}-f_l\right|^2  + \Sigma_{n,n}^{(i+1)} \right) + \sum_{n=1}^N \sum_{l=1}^L  g_{n,l}\widehat {\ln\gamma_n}^{(i+1)}    +   \sum_{n=1}^N \sum_{l=1}^L g_{n,l} \ln\rho_l,
\end{align}
where $\widehat {\ln\gamma_n}^{(i+1)} =\left< \ln \gamma_n \right>_{q^{(i+1)}(\gamma_n)}$.
Note that the assignment vector $\g_n$ only takes values from $\e_1, \e_2, \ldots, \e_L$, where the definition $\e_l$ is found in  Definition~1. Hence, we only have to calculate $q(\g_n=\e_l), l=1,2,\ldots, L$, to characterize the posterior distribution $q(\g_n)$, i.e.,
\begin{align}
&\ln q^{(i+1)} (\g_{n}=\e_l) \propto   \underbrace{\widehat{\ln \gamma_{n}}^{(i+1)} -
\hat{\gamma}_{n}^{(i+1)} \chi_{n,l}^{(i+1)}
+ \ln \rho_l
}_{\triangleq\nu_{n,l}^{(i+1)}}.
\end{align}
Since $ \sum_{l=1}^{L} q^{(i+1)} (\g_{n}=\e_l)=1$, we have
\begin{align}\label{eq-alg4}
\phi_{n,l}^{(i+1)}= q^{(i+1)} (\g_{n}=\e_l)  =  \frac{\exp(\nu_{n,l}^{(i+1)})}{\sum_{l=1}^L \exp(\nu_{n,l}^{(i+1)})}.
\end{align}

The proposed alternating optimization algorithm proceeds to repeatedly  updating (\ref{eq-alg1}), (\ref{eq-alg2}), (\ref{eq-alg3}) and (\ref{eq-alg4}) until it converges. We will discuss the initialization and  convergence property latter. Expressions used during the update can be  calculated as
\begin{align}
\hat{\alpha}^{(i+1)}=& \frac{ a+M }{b_\alpha^{(i+1)}},\label{eq-arpchi00}\\
\hat\gamma_n^{(i+1)}=& \frac{a+1}{ b_{n}^{(i+1)}},~~\forall n, \label{eq-arpchi11}\\
\widehat{\ln\gamma_{n}}^{(i+1)} =& \Psi \left( a+1 \right)  - \ln \left(b_{n}^{(i+1)}\right),~~\forall n,\label{eq-arpchi22}\\
\chi_{n,l}^{(i+1)}=& \left( \left|\mu_n^{(i+1)}-f_l\right|^2  + \Sigma_{n,n}^{(i+1)} \right),~~\forall n,l, \label{eq-arpchi}
\end{align}
where $\mu_{n}^{(i+1)}$ and $\Sigma_{n,n}^{(i+1)}$ stand for the $n$-th element and the $n$-th diagonal element of $\bm\mu^{(i+1)}$ and $\bm\Sigma^{(i+1)}$, respectively. Our VBI-based method for discrete signal reconstruction is outlined in Algorithm 1.
Note that the most demanding step in Algorithm 1 is to compute an inverse of an $N\times N$ matrix in Step~3-b.
To reduce the computational cost, we may adopt Woodbury matrix identity:
\begin{align}
&\bm\Sigma =  \bm\Delta -   \bm\Delta \A^H\left(\hat\alpha^{-1}\I_M + \A \bm\Delta \A^H \right)^{-1} \A  \bm\Delta,\label{invSi}
\end{align}
where $ \bm\Delta\triangleq (\sum_{l=1}^L \Q_l)^{-1}$ and the  iteration subscript is dropped for notational simplicity.
Finally, the main computational complexity per iteration is given as follows.
\begin{itemize}
  \item The complexity in updating  $q^{(i+1)}(\alpha)$ is $\mathcal{O}( M N^2  )$.
  \item The complexity in updating  $q^{(i+1)}(\x)$ is $\mathcal{O}(  M N^2 + M^3 )$, which can be simplified to $\mathcal{O}(  M N^2 )$ because of $N>M$.
  \item The complexity in updating  $q^{(i+1)}(\bm\gamma)$ is $\mathcal{O}( L N  )$.
  \item The complexity in updating  $q^{(i+1)}(\G)$ is $\mathcal{O}( L N  )$.
\end{itemize}
Therefore, the total computational complexity of Algorithm~1 is $\mathcal{O}(M N^2)$ per iteration.

\begin{algorithm}
\caption{{VBI-based Algorithm for Discrete Signal Reconstruction}}
\begin{enumerate}
  \item Input: $\y$, $\A$, $\mathcal{F}=\{f_l\}_{l=1}^L$ and  $\{\rho_l \}_{l=1}^L$.

  \item Initialization: Let $a=b=10^{-10}$ and $i=0$, and set $q^{(i)}(\x)$, $q^{(i)}(\bm\gamma)$ and $\phi^{(i)}_{n,l}, \forall n,l$, to initial values as in (\ref{eq-in1})--(\ref{eq-in3}).

     \item Repeat the following until it converges:
\begin{itemize}
     \item[a)] Update $q^{(i+1)}(\alpha)= \Gamma(\alpha| a + M , b_\alpha^{(i+1)}  )$ with (\ref{eq-alg1}), and calculate $\hat{\alpha}^{(i+1)}$ with (\ref{eq-arpchi00})

     \item[b)] Update $q^{(i+1)}(\x)= \mathcal{CN}(\x| \bm\mu^{(i+1)} ,  \bm\Sigma^{(i+1)}   )$ with (\ref{eq-alg2}), and calculate $\chi_{n,l}^{(i+1)}, \forall n,l$, with (\ref{eq-arpchi}).

     \item[c)] Update $q^{(i+1)}(\gamma_{n})= \Gamma\left(\gamma_{n} |a+1, b_{n}^{(i+1)}\right), \forall n$, with (\ref{eq-alg3}), and calculate $\hat\gamma_n^{(i+1)}$ and $\widehat{\ln\gamma_n}^{(i+1)}, \forall n$, with (\ref{eq-arpchi11}) and (\ref{eq-arpchi22}), respectively.

     \item[d)] Calculate $\phi_{n,l}^{(i+1)}, \forall n,l$, with (\ref{eq-alg4}).
     \item[e)] $i=i+1$.

\end{itemize}

  \item Output: $  x_n^{\mathrm{est}}= \arg\min_{f\in \mathcal{F}}| \mu_{n}^{(i)}  - f |^2, \forall n$.

\end{enumerate}
\end{algorithm}

\subsection{Initialization and  Convergence Analysis}

To start the alternating optimization algorithm, initialization for  $ q^{(0)}(\x)$, $q^{(0)}(\bm\gamma)$, and $q^{(0)}(\G)$ is needed.  According to (\ref{eq-alg2}), (\ref{eq-alg3}) and (\ref{eq-alg4}), these initial values are set as follows:
\begin{align}
q^{(0)}(\x)=&  \mathcal{CN}(\x|  (\A^H\A +\I_N )^{-1}\A^H\y   ,  (\A^H\A +\I_N )^{-1}    ),\label{eq-in1}\\
q^{(0)}(\bm\gamma)=& \prod_{n=1}^N \Gamma\left(\gamma_{n} |a+1,  b+1   )\right),\label{eq-in2}\\
\phi_{n,l}^{(0)}= & \frac{1}{L}, ~~\forall n,l.\label{eq-in3}
\end{align}
Empirical evidence illustrates that the proposed method is very robust to the above initialization.

In general, the convergence (to a stationary point) for an alternating algorithm cannot be guaranteed. However,  the alternating algorithm for our problem can be parameterized and reformulated as a special block majorization-minimization (MM) algorithm \cite{razaviyayn2014successive}, which enables us to prove that it converges to a stationary point as follows.

\noindent\textbf{Lemma~1.} If at each iteration, we do updates as in (\ref{eqM1})--(\ref{eqM4}), the generated iterates converge to a stationary point of the problem (\ref{eq-KLproblem}).
\begin{proof}
 See Appendix~A.
\end{proof}

\subsection{Challenge with  Ideal Prior (\ref{fa2})}

As mentioned in Remark~2, it is intractable to perform the Bayesian inference with  the ideal prior under the SBL framework, whose reason is given as follows. Replacing $p(x_n|\g_n,\gamma_n)$ by $p(x_n|\g_n)$, (\ref{eq-ut2}) can be rewritten as
\begin{align}
&\ln q^{(i+1)}(\x)\notag\\
\propto&  -\hat{\alpha}^{(i+1)}\| \y- \A\x \|_2^2  +\Big<\sum_{n=1}^N \ln\Big(\sum_{l=1}^L  g_{n,l} \delta(x_n-f_l)\Big)\Big>_{q^{(i)}(\G) }\notag\\
=& -\hat{\alpha}^{(i+1)}\| \y- \A\x \|_2^2, ~~ x_n\in \{f_1, f_2,\ldots, f_L\},\forall n, \label{eq-olqx}
\end{align}
where the last equality comes from the definition of $\delta(x_n-f_l)$.
Obviously, the feasible $\x$ can take values from $L^N$ candidates (denoted by $\c_1,\c_2, \ldots, \c_{L^N} $). If we can exhaustively calculate the value of $\ln q^{(i+1)}(\x=\c_j),\forall j$, the discrete distribution $q^{(i+1)}(\x)$ can be obtained similarly as in (\ref{eq-alg4}). Since the value of $N$ is usually large in discrete signal reconstruction problems, the massive computation involved in the exhaustive calculation could make the Bayesian inference
intractable for real applications.

In this paper, it is the first time to obtain a tractable SLB-based framework for discrete
signal reconstruction with the help of the new proposed discretization enforcing prior (\ref{eq-prix}).
Moreover, our VBI method does not impose any restrictions on the measurement matrix.
To the best of our knowledge, the SOAV optimization is the only method for discrete signal reconstruction with non i.i.d. Gaussian measurement matrices, but it has several shortcomings (see Section II-B). Our VBI method can overcome
all the shortcomings of the SOAV optimization, and simulation results illustrate that our VBI method can achieve the best performance for non i.i.d. Gaussian measurement matrices.

On the other hand, we may resort to the GAMP approximation \cite{rangan2010generalized} to overcome the challenge with ideal prior (\ref{fa2}).  In the next section, we will embed  GAMP into the VBI-based method to propose a fast GAMP variant, so as to directly adopt the ideal prior and significantly reduce the computational burden.
However, it is worth noting that the GAMP-based method works for i.i.d. Gaussian measurement matrices only.

\section{Fast GAMP for Discrete Signal Reconstruction}


In this section, we assume that the elements of the measurement matrix $\A$ are i.i.d.
Gaussian distributed. In this case, the GAMP algorithm \cite{rangan2010generalized} can be adopted to handle the  ideal discretization prior (\ref{eqideal0}). However,
the original GAMP algorithm needs the knowledge of the noise variance, which is usually unknown in practical scenarios. To jointly estimate the noise variance and reconstruct the unknown signal with the ideal prior, we embed GAMP into the proposed VBI-based method, which is inspired by the works in \cite{vila2013expectation,li2015computationally,yang2018fast,dai2021real}.  With adopting the ideal prior, the GAMP variant can achieve an excellent recovery for an i.i.d. Gaussian measurement matrix. Nevertheless, it might give a very bad performance when the i.i.d. Gaussian assumption is violated.

%
%
%


\subsection{GAMP Introduction}

GAMP is a low-complexity algorithm developed in a loopy belief-propagation framework for efficiently computing approximate marginal posteriors using the cental limit theorem. Since GAMP can deal with arbitrary distributions on both input and output, it can be applied to a wider range of CS problems.
Following the convention in GAMP, we introduce $z_m\triangleq\sum_{n=1}^N a_{mn}x_n,\forall m$, into (\ref{eq-csm}), i.e.,
\begin{align}\label{eq-cszz}
y_m=z_m + v_m,~~\forall m,
\end{align}
where $a_{mn}$ stands for the $(m,n)$ element of $\A$.
The original GAMP algorithm is outlined in Algorithm~2, where $\bm\Theta_m^{\mathrm{out}} $ and $\bm\Theta_n^{\mathrm{in}}$ stand for  the prior information about $v_m$ and $x_n$, respectively. We refer the reader to \cite{rangan2010generalized} for more details and background about GAMP.
Here, we do not introduce the messages among $x_n$s and $z_m$s and also do not illustrate how to derive Algorithm~2 with the message approximation, because all the messages and derivations remain unchanged when we combine the GAMP algorithm into the proposed VBI method, except that the following two distribution expressions are needed to be recalculated only:
\begin{itemize}
  \item In Steps 3-c and 3-d,  $p ( z_m| y_m,  \mu^{p}_m, \tau_m^p, \bm\Theta_m^{\mathrm{out}})$ corresponds to the approximation of the marginal posterior $p(z_m|y_m, \bm\Theta_m^{\mathrm{out}})$.
 \item In Steps 3-i and 3-j,  $p(x_n|\y, \mu^{r}_n, \tau_n^r, \bm\Theta_n^{\mathrm{in}})$ corresponds to the approximation of the marginal posterior $p(x_n|\y, \bm\Theta_n^{\mathrm{in}})$.
\end{itemize}
These two distribution functions are closely related to our proposed method, which will be utilized to approximately calculate (\ref{eq-alg1}) and (\ref{eq-alg2}), respectively.  We will detailedly discuss their calculations in the next subsection.

\begin{algorithm}
\caption{{GAMP Algorithm in \cite{rangan2010generalized}}}
\begin{enumerate}
  \item Input: $\y$, $\A$, $\bm\Theta_m^{\mathrm{out}} $ and $\bm\Theta_n^{\mathrm{in}}$, $\forall m,n$.

  \item Initialization: Set $t=1$ and set $T_{\mathrm{max}}$, $\mu_n^x$, $\tau_n^x$ and $\mu^{s}_m, \forall m,n$.

     \item Repeat the following until convergence or $t\le T_{\mathrm{max}}$:
\begin{itemize}
     \item[\%] \emph{Output Linear Step}
     \item[a)]  $\tau_m^p=\sum_{n=1}^N  |a_{mn}|^2 \tau_n^x$,~~$\forall m$.

     \item[b)]  $\mu^p_m= \sum_{n=1}^N a_{mn} \mu_n^x  -  \tau_m^p \mu^{s}_m$,~~$\forall m$.
     \item[\%] \emph{Output Nonlinear Step}
     \item[c)]  $\mu^{z}_m =\left< z_m \right>_{p ( z_m| y_m,  \mu^{p}_m, \tau_m^p, \bm\Theta_m^{\mathrm{out}} )}$,~~$\forall m$.
     \item[d)]   $\tau_m^z=\left< |z_m - \mu^{z}_m|^2  \right>_{p( z_m| y_m,  \mu^{p}_m, \tau_m^p, \bm\Theta_m^{\mathrm{out}} )}  $,~~$\forall m$.

     \item[e)]  $\mu^{s}_m=( \mu^{z}_m - \mu^{p}_m)/ \tau_m^p$,~~$\forall m$.
     \item[f)]  $\tau_m^s= ( 1-\tau_m^z/\tau_m^p)/\tau_m^p$,~~$\forall m$.
     \item[\%] \emph{Input Linear Step}
     \item[g)]  $\tau_n^r= \left( \sum_{m=1}^M |a_{mn}|^2 \tau_m^s  \right)^{-1} $,~~$\forall n$.
     \item[h)]  $\mu^{r}_n=\mu_n^x +\tau_n^r\sum_{m=1}^M a_{mn}\mu^{s}_m$,~~$\forall n$.
     \item[\%] \emph{Input Nonlinear Step}
     \item[i)]  $\mu_n^x=\left< x_n\right>_{p(x_n|\y, \mu^{r}_n, \tau_n^r, \bm\Theta_n^{\mathrm{in}}  )}$,~~$\forall n$.
     \item[j)]  $\tau_n^x= \left< |x_n  - \mu_n^x|^2 \right>_{p(x_n|\y, \mu^{r}_n, \tau_n^r, \bm\Theta_n^{\mathrm{in}}  )}$,~~$\forall n$.
     \item[k)]  $t=t+1$.

\end{itemize}

  \item Output: $\mu^{z}_m$, $\tau_m^z$, $\mu_n^x$ and $\tau_n^x$, $\forall n,m$.

\end{enumerate}
\end{algorithm}

\subsection{Approximation Details}

Note that
$\bm\Theta_m^{\mathrm{out}}=\{\hat\alpha\}$ and $\bm\Theta_n^{\mathrm{in}}=\{\bm\rho\triangleq \{\rho_{l}\}_{l=1}^{L} \}, \forall m,n$, in  our case, where the definition of $\rho_{n}$ has been given in (\ref{disg}). In the following, we recalculate the two approximate distributions $p ( z_m| y_m,  \mu^{p}_m, \tau_m^p, \bm\Theta_m^{\mathrm{out}})$ and $p(x_n|\y, \mu^{r}_n, \tau_n^r, \bm\Theta_n^{\mathrm{in}})$ one-by-one.

\begin{itemize}
  \item Firstly, according to EQ. (26) in \cite{rangan2010generalized}, the true marginal posterior $p(z_m|y_m, \hat\alpha)$ can be approximately calculated as:
\begin{align}
p( z_m| y_m,  \mu^{p}_m, \tau_m^p, \hat\alpha ) = \frac{p(y_m|z_m, \hat\alpha ) \mathcal{CN}(z_m| \mu^{p}_m, \tau_m^p)  }{\int_z  p(y_m|z, \hat\alpha) \mathcal{CN}(z| \mu^{p}_m, \tau_m^p) dz  } ,\label{eqGAMPpro1}
\end{align}
where $\mu^{p}_m$ and  $\tau_m^p$ vary with the GAMP iteration $t$ (as shown in Steps 3-b and 3-a). Here, the iteration index is dropped for simplicity.
Under the assumption of the additive complex i.i.d. Gaussian noises, we have $p(y_m|z_m,\hat\alpha)=  \mathcal{CN} (y_m| z_m, \hat\alpha^{-1} )$. Therefore, $p( z_m| y_m,  \mu^{p}_m, \tau_m^p, \hat\alpha ) $ obeys a complex Gaussian distribution:
\begin{align}
p ( z_m| y_m,  \mu^{p}_m, \tau_m^p, \hat\alpha )
=  \mathcal{CN} ( z_m|   \mu_m^z, \tau_m^z   ) , \label{apro1}
\end{align}
where
\begin{align}
\mu_m^z=&\frac{\hat\alpha\tau_m^p y_m + \mu^{p}_m  }{1+ \hat\alpha\tau_m^p},\\
\tau_m^z=& \frac{\tau_m^p }{1+ \hat\alpha\tau_m^p}.
\end{align}

  \item
Secondly, according to EQ. (19) in \cite{rangan2010generalized}, $p(x_n|\y, \bm\rho) $ can be approximately calculated as:
  \begin{align}
  &p(x_n|\y, \mu^{r}_n, \tau_n^r, \bm\rho )\notag\\
  =& \frac{  p(x_n |\bm\rho)   \mathcal{CN}(x_n| \mu^{r}_n, \tau_n^r)  }{\int_x p(x| \bm\rho ) \mathcal{CN}(x| \mu^{r}_n, \tau_n^r) dx  }\\
  =& \frac{ \mathcal{CN}(x_n| \mu^{r}_n, \tau_n^r)\cdot \sum_{\g_n\in \{\e_l  \}_{l=1}^L } p(x_n| \g_n )  p( \g_n )}
  {\int_x p(x| \bm\rho ) \mathcal{CN}(x| \mu^{r}_n, \tau_n^r) dx},  \label{eqGAMPpro2}
  \end{align}
  where $\mu^{r}_n$ and  $\tau_n^r$ will be again  updated in every iteration of GAMP (as shown in Steps 3-h and 3-g). Substituting  (\ref{fa2}) and (\ref{disg2}) into (\ref{eqGAMPpro2}) results in
  \begin{align}
  &p(x_n|\y, \mu^{r}_n, \tau_n^r, \bm\rho )= \frac{ \mathcal{CN}(x_n| \mu^{r}_n, \tau_n^r) \cdot \sum_{l=1}^L \rho_l \delta(x_n-f_l) }{\int_x p(x|\bm\rho) \mathcal{CN}(x| \mu^{r}_n, \tau_n^r) dx  }. \label{apro2}
  \end{align}
  Clearly, $p(x_n|\y, \mu^{r}_n, \tau_n^r, \bm\rho )$ is a  discrete distribution which only takes values from the  finite alphabet $\mathcal{F}$ with the probabilities
  \begin{align}\label{eq-pxnl}
   p^x_{nl}  = \frac{\rho_l}{c_n}  \exp\left(- \frac{|f_l-  \mu^{r}_n|^2 }{\tau_n^r} \right), ~~\forall n, l,
  \end{align}
  where $p^x_{nl}$ is short for $p(x_n=f_l|\y, \mu^{r}_n, \tau_n^r, \bm\rho )$ and
  $c_n = \pi |\tau_n^r| \cdot  \int_x p(x|\bm\rho) \mathcal{CN}(x| \mu^{r}_n, \tau_n^r) dx $ is
  a constant. Since $\sum_{l=1}^L p^x_{nl} =1$, $c_n$ can be alternatively calculated as
  \begin{align}
   c_n = \sum_{l=1}^L\rho_l  \exp\left(- \frac{|f_l-  \mu^{r}_n|^2 }{\tau_n^r} \right).
  \end{align}
Based on the definitions of $\mu_{n}^x$ and $\tau_n^x$ in Steps 3-i) and 3-j), we have
\begin{align}
\mu_{n}^x= &\sum_{l=1}^L f_l p^x_{nl} ,  \\
\tau_n^x=&  \sum_{l=1}^L | f_l - \mu_{n}^x  |^2 p^x_{nl} .
\end{align}
\end{itemize}

It is seen from (\ref{apro2}) that the GAMP-based method separates $p(\x|\y, \bm\rho)$ into $N$ independent discrete marginal posteriors approximately (i.e., $p(\x|\y,\bm\rho)\approx \prod_{n=1}^N p(x_n|\y, \mu^{r}_n, \tau_n^r, \bm\rho )$). Such separation can reduce the number of the total discrete candidates from $L^N$ to $NL$. Hence, it is tractable to calculate the  discrete distribution $p(\x|\y,\bm\rho)$ approximately with the GAMP-based method.

\subsection{Propposed GAMP-based Extension}

Recall that our method proposed in Section III only has to repeatedly  update (\ref{eq-alg1}), (\ref{eq-alg2}), (\ref{eq-alg3}) and (\ref{eq-alg4}). In the following, we illustrate how to embed the approximations (\ref{apro1}) and (\ref{apro2}) into these updates. For ease of notation, the iteration subscript is dropped in this subsection.

\subsubsection{Approximation for (\ref{eq-alg1})} In order to combine the approximation (\ref{apro1}) with (\ref{eq-alg1}), we rewrite
(\ref{eq-upal01}) as
\begin{align}
\ln q(\alpha)
\propto& (a + M -1)\ln \alpha -  \alpha\cdot\left( b+   \sum_{m=1}^M \left< |y_m - z_m |^2 \right>_{p ( z_m| y_m,  \mu^{p}_m, \tau_m^p, \alpha )} \right) \label{eq-GAMPupal1}\\
 \propto& (a + M -1)\ln \alpha -  \alpha\cdot\left( b+  \sum_{m=1}^M  \left(| y_m - \mu^z_m |^2     +  \tau_m^z  \right)   \right).
\end{align}
Hence, we obtain
\begin{align}\label{eq-Galg1}
q(\alpha)\approx& \Gamma(\alpha| a + M , \sum_{m=1}^M  (| y_m - \mu^z_m |^2  +  \tau_m^z  ) )
\end{align}
and
\begin{align}
\hat\alpha=\left<\alpha\right>_{q(\alpha)}\approx& \frac{ a+M }{\sum_{m=1}^M  (| y_m - \mu^z_m |^2  +  \tau_m^z  )}.\label{eq-Garpchi00}
\end{align}

\subsubsection{Approximation for (\ref{eq-alg2})}  We approximate $q(x_n)$  by the discrete distribution $p(x_n|\y, \mu^{r}_n, \tau_n^r, \bm\rho)$, where $x_n$ only takes values from the  finite alphabet $\mathcal{F}$ with the probabilities
$p^x_{n,l}$ defined in (\ref{eq-pxnl}). Note that $p^x_{n,l}$s fully indicate the alphabet $x_n$ should take.

Obviously, neither $\bm\gamma$ nor $\G$ is required for updating $q(\alpha)$ and $q(x_n)$s.
Once $q(x_n)$ are obtained, the final discrete value of $x_n$ can be determined by  the maximum element of $\{p^x_{n,1}, p^x_{n,2},\ldots, p^x_{n,L} \}$.
Therefore,  the updates (\ref{eq-alg3}) and (\ref{eq-alg4}) can be safely removed from the fast GAMP-based method.
Empirical evidence shows that it remains very robust to the above GAMP approximations. We can always set $T_{\mathrm{max}}=1$ when Algorithm~2 is evoked, which means just one iteration is sufficient for the GAMP approximation. The proposed fast GAMP-based algorithm for discrete signal reconstruction is outlined in Algorithm~3.

\begin{algorithm}
\caption{{Fast GAMP-based Algorithm for Discrete Signal Reconstruction}}
\begin{enumerate}
 \item Input: $\y$, $\A$, $\mathcal{F}=\{f_l\}_{l=1}^L$ and  $\{\rho_l \}_{l=1}^L$.

  \item Initialization:  Set $\mu_n^x=[(\A^H\A +\I )^{-1}\A^H\y ]_n$, $\tau_n^x=1$ and $\mu^{s}_m=0$, $\forall m,n$, and let  $a=b=10^{-10}$ and  $\mu^z_m=\sum_{n=1}^N a_{mn}\mu^x_n$.

     \item Repeat the following until convergence:
\begin{itemize}
     \item[a)] Approximate $q(\alpha)$ using (\ref{eq-Galg1}), and calculate $\hat{\alpha}$ with (\ref{eq-Garpchi00}).

     \item[b)] Approximate $q(x_n)$, $\forall n$, by evoking the GAMP approximation:
                 \begin{itemize}
                 \item  $\tau_m^p=\sum_{n=1}^N  |a_{mn}|^2 \tau_n^x$,~~$\forall m$.

                 \item  $\mu^p_m= \sum_{n=1}^N a_{mn} \mu_n^x  -  \tau_m^p \mu^{s}_m$,~~$\forall m$.
                 \item  $\mu^{z}_m =(\hat{\alpha}\tau_m^p y_m + \mu^{p}_m )/(1+ \hat{\alpha}\tau_m^p)$,~~$\forall m$.
                 \item   $\tau^{z}_m=\tau_m^p/(1+ \hat\alpha\tau_m^p)$,~~$\forall m$.

                 \item  $\mu^{s}_m=( \mu^{z}_m - \mu^{p}_m)/ \tau_m^p$,~~$\forall m$.
                 \item $\tau_m^s= ( 1-\tau_m^z/\tau_m^p)/\tau_m^p$,~~$\forall m$.
                 \item  $\tau_n^r= \left( \sum_{m=1}^M |a_{mn}|^2 \tau_m^s  \right)^{-1} $,~~$\forall n$.
                 \item  $\mu^{r}_n=\mu_n^x +\tau_n^r\sum_{m=1}^M a_{mn}\mu^{s}_m$,~~$\forall n$.

                 \item

                  $p^x_{nl}  = \frac{\rho_l\exp\left(- \frac{|f_l-  \mu^{r}_n|^2 }{\tau_n^r} \right)}{\sum_{l=1}^L\rho_l  \exp\left(- \frac{|f_l-  \mu^{r}_n|^2 }{\tau_n^r} \right)}$,~~$\forall n,l$.
                 \item  $\mu_n^x=\sum_{l=1}^L f_l p^x_{nl}$,~~$\forall n$.
                 \item  $\tau_n^x= \sum_{l=1}^L | f_l - \mu_{n}^x  |^2 p^x_{nl}$,~~$\forall n$.
         \end{itemize}



%

\end{itemize}

  \item Output: $  x_n^{\mathrm{est}}= \arg \min_{f\in \mathcal{F}}| \mu_{n}^x  - f   |^2, \forall n$.

\end{enumerate}
\end{algorithm}

Finally, the main computational burden of Algorithm~3 is given as follows.
\begin{itemize}
  \item The complexity in Step~3-a is $O(M)$ per iteration.
  \item The complexity in Step~3-b is $O(MN)$ per iteration.
\end{itemize}
Therefore, the total computational complexity of  Algorithm~3 is $\mathcal{O}(MN)$ per iteration, which is much less than $\mathcal{O}(MN^2)$ for Algorithm~1. Simulation results in Section V will illustrate that
the GAMP-based method can achieve an excellent recovery for an i.i.d. Gaussian measurement matrix because the ideal prior (\ref{fa2}) is exploited, but its performance will degrade substantially for a non i.i.d. Gaussian $\A$.

\section{Simulation Results}

In this section, we  present simulation results to illustrate the performance of our method, with comparison to the following schemes:
\begin{itemize}
\item \textbf{Baseline~1} (Original SOAV): The discrete signal  is recovered using the original SOAV method \cite{nagahara2015discrete}.
\item \textbf{Baseline~2} (Optimal SOAV): The discrete signal  is recovered using the optimal SOAV method proposed in Section IV of \cite{hayakawa2018discreteness}.
\item \textbf{Baseline~3} (BODAMP): The discrete signal  is recovered using the Bayes optimal discreteness-aware AMP method proposed in Section V of \cite{hayakawa2018discreteness}.

\item \textbf{Baseline~4} (Standard SBL): $\x$  is recovered by using the standard SBL method \cite{tipping2001sparse} and the discrete signal is obtained by  projecting $\x$ onto the discrete set $\mathcal{F}$.

\end{itemize}

Two types of measurement matrices will be used: 1) i.i.d. Gaussian measurement matrix and 2) correlated measurement matrix. For i.i.d. Gaussian measurement matrix $\A$, it has i.i.d. zero-mean circularly symmetric complex Gaussian entries with variance $1/M$; while for a correlated measurement matrix $\A$, it is in the form of $\A= \R_M^{\frac{1}{2}} \A_{\mathrm{iid}} \R_N^{\frac{1}{2}} $ \cite{hayakawa2018discreteness,shin2003capacity}, where $\R_M$ (or $\R_N$) stands for an $M\times M$ (or $N\times N$) positive definite matrix with $(i,j)$ element being $J_0(|i-j|\pi)$ and $J_0(\cdot)$ stands for the zeroth-order Bessel function of the first kind.
Unless otherwise specified, in the following, we assume that the $L$ elements of the finite alphabet $\mathcal{F}$ are uniformly located on the unit circle in the complex plane, and the corresponding probabilities $\rho_l$s are randomly chosen with a uniform distribution.
All the simulations are conducted on an Intel Core i5-11400 CPU with 32 GB RAM using MATLAB R2020b.

\subsection{MSE Performance Versus Iteration Number}

\begin{figure}
\center
\begin{tikzpicture}[scale=1]
\begin{semilogyaxis}
[
ylabel={MSE}, grid=both,
legend style={at={(0.52,0.58),
font=\footnotesize},
anchor=north,legend columns=1
}, xmin=1,ymin=0.0001,xmax=80,title={(a)},width=10cm,height=5.88cm]
\addplot[mark=asterisk,red] file {Fig1/a/SBL.txt} ;    
\addplot[color=blue,mark=o] file {Fig1/a/GAMP.txt} ;
\addplot[color=black,mark=triangle] file {Fig1/a/BODAMP.txt} ;
\addplot[color=black,mark=+] file {Fig1/a/SOAV.txt} ;
\addplot[color=black,mark=diamond] file {Fig1/a/Original_SOAV.txt} ;
\addplot[color=black,mark=square] file {Fig1/a/Original_SBL.txt} ;
\legend{Proposed VBI, Proposed GAMP, BODAMP, Optimal SOAV, Original SOAV, Standard SBL}
\end{semilogyaxis}
\end{tikzpicture}
~
\begin{tikzpicture}[scale=1]
\begin{semilogyaxis}
[xlabel={Number of iterations},
ylabel={MSE}, grid=both,
legend style={at={(0.58,0.43),
font=\footnotesize},
anchor=north,legend columns=1
}, xmin=1,ymin=0.0001,xmax=80,title={(b)},width=10cm,height=5.88cm]
\addplot[mark=asterisk,red] file {Fig1/ainf/SBL.txt} ;    
\addplot[color=blue,mark=o] file {Fig1/ainf/GAMP.txt} ;
\addplot[color=black,mark=triangle] file {Fig1/ainf/BODAMP.txt} ;
\addplot[color=black,mark=+] file {Fig1/ainf/SOAV.txt} ;
\addplot[color=black,mark=diamond] file {Fig1/ainf/Original_SOAV.txt} ;
\addplot[color=black,mark=square] file {Fig1/ainf/Original_SBL.txt} ;
\end{semilogyaxis}
\end{tikzpicture}
\caption{MSE  versus the number of iterations with an i.i.d. Gaussian measurement matrix and $N=100$. a) $\Delta=0.7$, $L=8$ and  SNR$=30$ dB;  b) $\Delta=0.8$, $L=16$ and  noise-free.
}\label{fig-converge-G}
\end{figure}
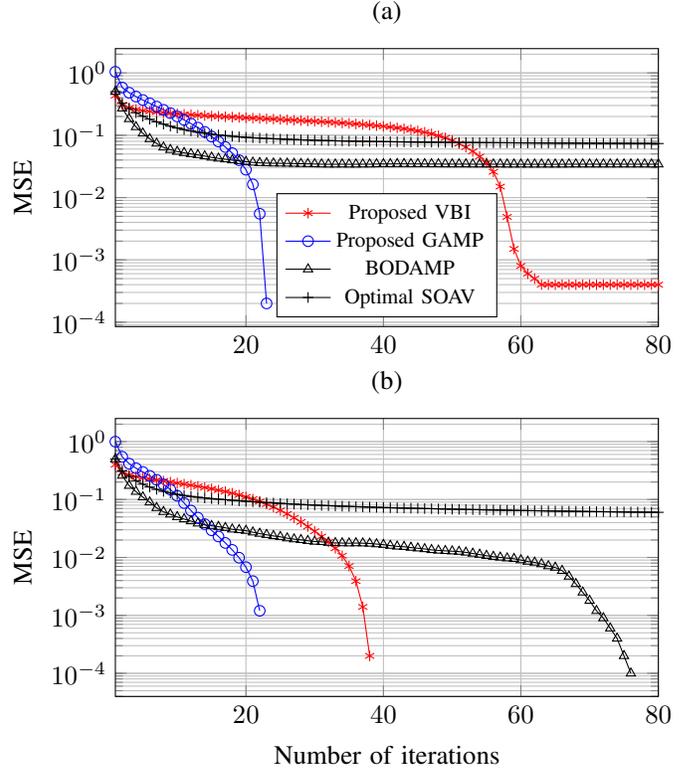

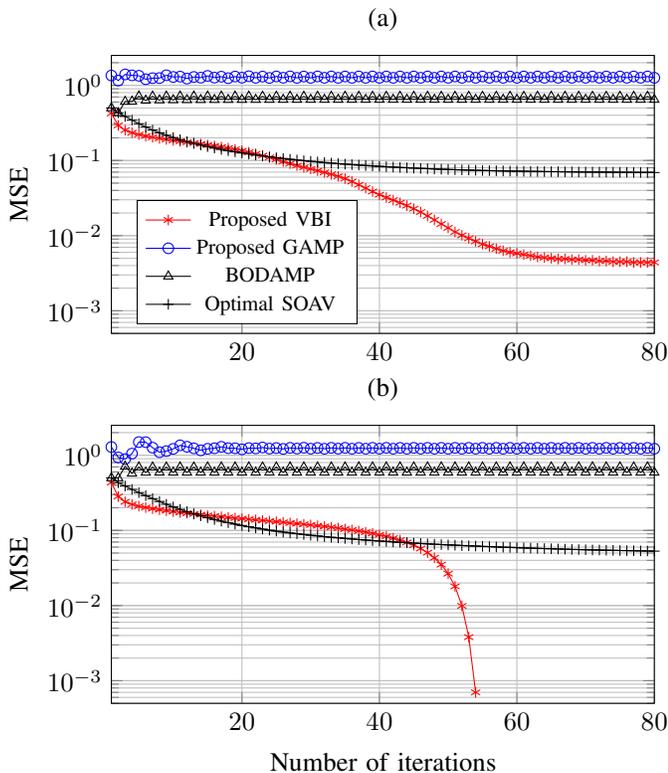
\begin{figure}
\center
\begin{tikzpicture}[scale=1]
\begin{semilogyaxis}
[
ylabel={MSE},grid=both,
legend style={at={(0.195,0.62),
font=\footnotesize},
anchor=north,legend columns=1
}, xmin=1,ymin=0,xmax=80,ymin=0.0001,ymax=2.5,title={(a)},width=10cm,height=5.88cm]
\addplot[mark=asterisk,red] file {Fig1/b820/SBL.txt} ;    
\addplot[color=blue,mark=o] file {Fig1/b820/GAMP.txt} ;
\addplot[color=black,mark=triangle] file {Fig1/b820/BODAMP.txt} ;
\addplot[color=black,mark=+] file {Fig1/b820/SOAV.txt} ;
\addplot[color=black,mark=diamond] file {Fig1/b820/Original_SOAV.txt} ;
\addplot[color=black,mark=square] file {Fig1/b820/Original_SBL.txt} ;
\legend{Proposed VBI, Proposed GAMP, BODAMP, Optimal SOAV, Original SOAV, Standard SBL}
\end{semilogyaxis}
\end{tikzpicture}
~
\begin{tikzpicture}[scale=1]
\begin{semilogyaxis}
[xlabel={Number of iterations},
ylabel={MSE}, grid=both,
legend style={at={(0.72,0.43),
font=\footnotesize},
anchor=north,legend columns=1
}, xmin=1,ymin=0,xmax=80,ymin=0.0001,ymax=2.5,title={(b)},width=10cm,height=5.88cm]
\addplot[mark=asterisk,red] file {Fig1/binf/SBL.txt} ;    
\addplot[color=blue,mark=o] file {Fig1/binf/GAMP.txt} ;
\addplot[color=black,mark=triangle] file {Fig1/binf/BODAMP.txt} ;
\addplot[color=black,mark=+] file {Fig1/binf/SOAV.txt} ;
\addplot[color=black,mark=diamond] file {Fig1/binf/Original_SOAV.txt} ;
\addplot[color=black,mark=square] file {Fig1/binf/Original_SBL.txt} ;
\end{semilogyaxis}
\end{tikzpicture}
\caption{MSE  versus the number of iterations with a correlated measurement matrix and $N=100$. a) $\Delta=0.7$, $L=8$ and  SNR$=30$ dB;  b) $\Delta=0.8$, $L=16$ and  noise-free.
}\label{fig-converge-C}
\end{figure}

In Figs.~\ref{fig-converge-G} and \ref{fig-converge-C}, we study the convergence and mean square error (MSE) performance for different discrete signal reconstruction strategies.
Let $N=100$ and $\Delta=M/N$, and the MSE at the $i$-th iteration is defined as \begin{align}
\mathrm{MSE}^{(i)}=\frac{\| \x^{(i)}_\mu-\x^{\mathrm{true}}\|_2^2}{N}
\end{align}
with $\x^{(i)}_\mu$ being the estimate of the true signal $\x^{\mathrm{true}}$ at the $i$-th iteration (without a hard decision). Fig.~\ref{fig-converge-G} shows
the MSE performance of the discrete signal reconstruction achieved by the different strategies with an i.i.d. Gaussian measurement matrix versus the number of iterations; while  Fig.~\ref{fig-converge-C} shows
the MSE performance of the discrete signal reconstruction achieved by the different strategies with a correlated measurement matrix versus the number of iterations.
It is observed that
(i) the GAMP-based method can yield the minimum MSE with an i.i.d. Gaussian measurement matrix (see Figs.~\ref{fig-converge-G}a and \ref{fig-converge-G}b), as well as the fastest convergence, because it can adopt the ideal prior (\ref{fa2}) directly; (ii) the GAMP-based method fails to  work with a correlated measurement matrix (see Figs.~\ref{fig-converge-C}a and \ref{fig-converge-C}b), as the GAMP approximation is designed for an i.i.d, Gaussian measurement matrix only; (iii) the VBI-based method works well for either an  i.i.d. Gaussian measurement matrix or a correlated measurement matrix; (iv)  the VBI-based method can achieve very similar performance in the noise-free case (see Figs.~\ref{fig-converge-G}b and \ref{fig-converge-C}b); (v) the SOAV-type methods outperforms BODAMP with a correlated measurement matrix (see Figs.~\ref{fig-converge-C}a and \ref{fig-converge-C}b), because the AMP-based method (BODAMP) also relies on the i.i.d. Gaussian assumption; (vi) the standard SBL method always fails to work as it cannot handle the discrete signal;
(vii) the VBI-based method has much smaller MSE than the SOAV-type method, no matter what the measurement matrix is used; and (viii) the VBI-based method may require more iteration numbers in some cases, but it almost converges within 70 iterations.


\subsection{SER Performance Versus SNR}

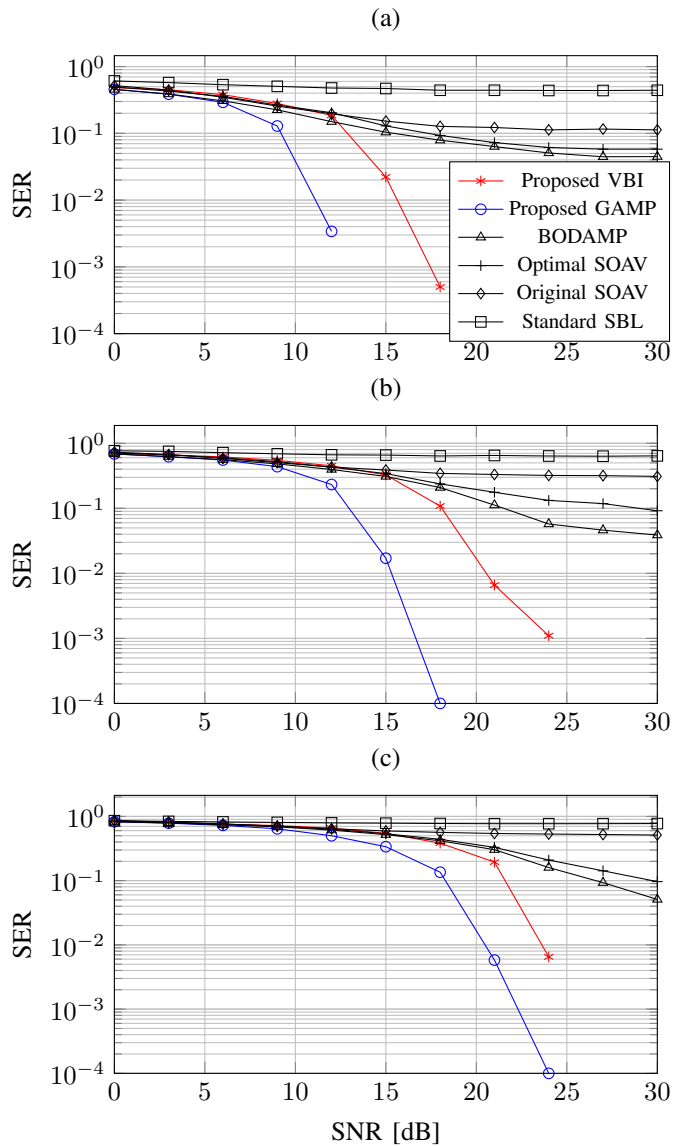
\begin{figure}
\center
\begin{tikzpicture}[scale=1]
\begin{semilogyaxis}[
ylabel={SER},grid=both, title={(a)},
legend style={at={(0.805,0.575),font=\footnotesize},
anchor=north,legend columns=1}, xmin=0,xmax=30,ymin=0.0001,width=10cm,height=5.88cm]
\addplot[mark=asterisk,red]  coordinates{
  (    0.0000 ,      0.4852)
  (    3.0000 ,      0.4507)
  (    6.0000 ,      0.3790)
  (    9.0000 ,      0.2799)
  (   12.0000 ,      0.1858)
  (   15.0000 ,      0.0222)
  (   18.0000 ,      0.0005)
  (   21.0000 ,           0)
  (   24.0000 ,           0)
  (   27.0000 ,           0)
  (   30.0000 ,           0)
};
\addplot[mark=o,blue]  coordinates{
  (    0.0000 ,       0.4566)
  (    3.0000 ,       0.3857)
  (    6.0000 ,       0.2913)
  (    9.0000 ,       0.1287)
  (   12.0000 ,       0.0034)
  (   15.0000 ,            0)
  (   18.0000 ,            0)
  (   21.0000 ,            0)
  (   24.0000 ,            0)
  (   27.0000 ,            0)
  (   30.0000 ,            0)
};
\addplot[mark=triangle]  coordinates{
  (    0.0000 ,      0.4546)
  (    3.0000 ,      0.3878)
  (    6.0000 ,      0.3066)
  (    9.0000 ,      0.2248)
  (   12.0000 ,      0.1497)
  (   15.0000 ,      0.1040)
  (   18.0000 ,      0.0789)
  (   21.0000 ,      0.0634)
  (   24.0000 ,      0.0508)
  (   27.0000 ,      0.0446)
  (   30.0000 ,      0.0448)
};
\addplot[mark=+] coordinates{
  (    0.0000 ,      0.4932)
  (    3.0000 ,      0.4295)
  (    6.0000 ,      0.3576)
  (    9.0000 ,      0.2649)
  (   12.0000 ,      0.2037)
  (   15.0000 ,      0.1291)
  (   18.0000 ,      0.0930)
  (   21.0000 ,      0.0727)
  (   24.0000 ,      0.0614)
  (   27.0000 ,      0.0579)
  (   30.0000 ,      0.0579)
};
\addplot[mark=diamond] coordinates{
  (    0.0000 ,       0.5193)
  (    3.0000 ,       0.4438)
  (    6.0000 ,       0.3474)
  (    9.0000 ,       0.2557)
  (   12.0000 ,       0.1966)
  (   15.0000 ,       0.1524)
  (   18.0000 ,       0.1271)
  (   21.0000 ,       0.1223)
  (   24.0000 ,       0.1126)
  (   27.0000 ,       0.1158)
  (   30.0000 ,       0.1127)
};
\addplot[mark=square] coordinates{
  (    0.0000 ,       0.6096)
  (    3.0000 ,       0.5751)
  (    6.0000 ,       0.5355)
  (    9.0000 ,       0.5056)
  (   12.0000 ,       0.4772)
  (   15.0000 ,       0.4694)
  (   18.0000 ,       0.4419)
  (   21.0000 ,       0.4419)
  (   24.0000 ,       0.4385)
  (   27.0000 ,       0.4380)
  (   30.0000 ,       0.4418)
};
\legend{Proposed VBI, Proposed GAMP, BODAMP, Optimal SOAV, Original SOAV, Standard SBL}
\end{semilogyaxis}
\end{tikzpicture}
~
\begin{tikzpicture}[scale=1]
\begin{semilogyaxis}[
ylabel={SER},grid=both, title={(b)},
legend style={at={(0.25,0.52),font=\footnotesize},
anchor=north,legend columns=1}, xmin=0,xmax=30,ymin=0.0001,width=10cm,height=5.88cm]
\addplot[mark=asterisk,red]  coordinates{
  (    0.0000 ,   0.6986  )
  (    3.0000 ,   0.6640  )
  (    6.0000 ,   0.6180  )
  (    9.0000 ,   0.5471  )
  (   12.0000 ,   0.4509  )
  (   15.0000 ,   0.3262  )
  (   18.0000 ,   0.1086  )
  (   21.0000 ,   0.0066  )
  (   24.0000 ,   0.0011  )
  (   27.0000 ,        0  )
  (   30.0000 ,        0  )
};
\addplot[mark=o,blue]  coordinates{
  (    0.0000 ,    0.6855 )
  (    3.0000 ,    0.6215 )
  (    6.0000 ,    0.5527 )
  (    9.0000 ,    0.4373 )
  (   12.0000 ,    0.2323 )
  (   15.0000 ,    0.0171 )
  (   18.0000 ,    0.0001 )
  (   21.0000 ,         0 )
  (   24.0000 ,         0 )
  (   27.0000 ,         0 )
  (   30.0000 ,         0 )
};
\addplot[mark=triangle]  coordinates{
  (    0.0000 ,      0.6866 )
  (    3.0000 ,      0.6238 )
  (    6.0000 ,      0.5607 )
  (    9.0000 ,      0.4823 )
  (   12.0000 ,      0.3957 )
  (   15.0000 ,      0.3093 )
  (   18.0000 ,      0.2087 )
  (   21.0000 ,      0.1121 )
  (   24.0000 ,      0.0574 )
  (   27.0000 ,      0.0462 )
  (   30.0000 ,      0.0388 )
};
\addplot[mark=+] coordinates{
  (    0.0000 ,      0.7106)
  (    3.0000 ,      0.6635)
  (    6.0000 ,      0.5970)
  (    9.0000 ,      0.5220)
  (   12.0000 ,      0.4293)
  (   15.0000 ,      0.3421)
  (   18.0000 ,      0.2377)
  (   21.0000 ,      0.1764)
  (   24.0000 ,      0.1324)
  (   27.0000 ,      0.1183)
  (   30.0000 ,      0.0918)
};
\addplot[mark=diamond] coordinates{
  (    0.0000 ,      0.7311)
  (    3.0000 ,      0.6715)
  (    6.0000 ,      0.5866)
  (    9.0000 ,      0.5043)
  (   12.0000 ,      0.4315)
  (   15.0000 ,      0.3880)
  (   18.0000 ,      0.3446)
  (   21.0000 ,      0.3316)
  (   24.0000 ,      0.3205)
  (   27.0000 ,      0.3181)
  (   30.0000 ,      0.3099)
};
\addplot[mark=square] coordinates{
  (    0.0000 ,      0.7684)
  (    3.0000 ,      0.7506)
  (    6.0000 ,      0.7179)
  (    9.0000 ,      0.6892)
  (   12.0000 ,      0.6653)
  (   15.0000 ,      0.6589)
  (   18.0000 ,      0.6387)
  (   21.0000 ,      0.6500)
  (   24.0000 ,      0.6378)
  (   27.0000 ,      0.6332)
  (   30.0000 ,      0.6406)
};
\end{semilogyaxis}
\end{tikzpicture}
~
\begin{tikzpicture}[scale=1]
\begin{semilogyaxis}[xlabel={SNR [dB]},
ylabel={SER},grid=both, title={(c)},
legend style={at={(0.25,0.52),font=\footnotesize},
anchor=north,legend columns=1}, xmin=0,xmax=30,ymin=0.0001,width=10cm,height=5.88cm]
\addplot[mark=asterisk,red]  coordinates{
  (    0.0000 ,      0.8520)
  (    3.0000 ,      0.8041)
  (    6.0000 ,      0.7568)
  (    9.0000 ,      0.7204)
  (   12.0000 ,      0.6571)
  (   15.0000 ,      0.5455)
  (   18.0000 ,      0.3830)
  (   21.0000 ,      0.1944)
  (   24.0000 ,      0.0065)
  (   27.0000 ,           0)
  (   30.0000 ,           0)
};
\addplot[mark=o,blue]  coordinates{
  (    0.0000 ,       0.8260)
  (    3.0000 ,       0.7837)
  (    6.0000 ,       0.7251)
  (    9.0000 ,       0.6410)
  (   12.0000 ,       0.5001)
  (   15.0000 ,       0.3373)
  (   18.0000 ,       0.1353)
  (   21.0000 ,       0.0058)
  (   24.0000 ,       0.0001)
  (   27.0000 ,            0)
  (   30.0000 ,            0)
};
\addplot[mark=triangle]  coordinates{
  (    0.0000 ,      0.8269)
  (    3.0000 ,      0.7900)
  (    6.0000 ,      0.7426)
  (    9.0000 ,      0.6864)
  (   12.0000 ,      0.6117)
  (   15.0000 ,      0.5198)
  (   18.0000 ,      0.4172)
  (   21.0000 ,      0.3035)
  (   24.0000 ,      0.1599)
  (   27.0000 ,      0.0930)
  (   30.0000 ,      0.0510)
};
\addplot[mark=+] coordinates{
  (    0.0000 ,       0.8431)
  (    3.0000 ,       0.8071)
  (    6.0000 ,       0.7607)
  (    9.0000 ,       0.7073)
  (   12.0000 ,       0.6308)
  (   15.0000 ,       0.5369)
  (   18.0000 ,       0.4399)
  (   21.0000 ,       0.3296)
  (   24.0000 ,       0.2091)
  (   27.0000 ,       0.1422)
  (   30.0000 ,       0.0966)
};
\addplot[mark=diamond] coordinates{
  (    0.0000 ,       0.8588)
  (    3.0000 ,       0.8146)
  (    6.0000 ,       0.7607)
  (    9.0000 ,       0.7050)
  (   12.0000 ,       0.6426)
  (   15.0000 ,       0.5939)
  (   18.0000 ,       0.5636)
  (   21.0000 ,       0.5421)
  (   24.0000 ,       0.5311)
  (   27.0000 ,       0.5223)
  (   30.0000 ,       0.5127)
};
\addplot[mark=square] coordinates{
  (    0.0000 ,       0.8575)
  (    3.0000 ,       0.8365)
  (    6.0000 ,       0.8195)
  (    9.0000 ,       0.8066)
  (   12.0000 ,       0.7923)
  (   15.0000 ,       0.7844)
  (   18.0000 ,       0.7752)
  (   21.0000 ,       0.7698)
  (   24.0000 ,       0.7696)
  (   27.0000 ,       0.7707)
  (   30.0000 ,       0.7740)
};
\end{semilogyaxis}
\end{tikzpicture}
\caption{SER  versus SNR with an i.i.d. Gaussian measurement matrix and $N=100$. a) $\Delta=0.7$ and $L=4$; b)  $\Delta=0.8$ and $L=8$; c) $\Delta=0.9$ and $L=16$.
 }\label{fig-SNR-G}
\end{figure}

\begin{figure}
\center
\begin{tikzpicture}[scale=1]
\begin{semilogyaxis}[
ylabel={SER},grid=both, title={(a)},
legend style={at={(0.195,0.66),font=\footnotesize},
anchor=north,legend columns=1}, xmin=0,xmax=30,ymin=0.0004,width=10cm,height=5.88cm]
\addplot[mark=asterisk,red]  coordinates{
  (    0.0000 ,      0.5108)
  (    3.0000 ,      0.4699)
  (    6.0000 ,      0.4226)
  (    9.0000 ,      0.3352)
  (   12.0000 ,      0.2505)
  (   15.0000 ,      0.0969)
  (   18.0000 ,      0.0040)
  (   21.0000 ,           0)
  (   24.0000 ,           0)
  (   27.0000 ,           0)
  (   30.0000 ,           0)
};
\addplot[mark=o,blue]  coordinates{
  (    0.0000 ,      0.6327)
  (    3.0000 ,      0.6354)
  (    6.0000 ,      0.6403)
  (    9.0000 ,      0.6469)
  (   12.0000 ,      0.6469)
  (   15.0000 ,      0.6476)
  (   18.0000 ,      0.6395)
  (   21.0000 ,      0.6431)
  (   24.0000 ,      0.6264)
  (   27.0000 ,      0.6474)
  (   30.0000 ,      0.6461)
};
\addplot[mark=triangle]  coordinates{
  (    0.0000 ,     0.6809)
  (    3.0000 ,     0.6822)
  (    6.0000 ,     0.6803)
  (    9.0000 ,     0.6828)
  (   12.0000 ,     0.6794)
  (   15.0000 ,     0.6873)
  (   18.0000 ,     0.6832)
  (   21.0000 ,     0.6880)
  (   24.0000 ,     0.6856)
  (   27.0000 ,     0.6841)
  (   30.0000 ,     0.6783)
};
\addplot[mark=+] coordinates{
  (    0.0000 ,      0.5212)
  (    3.0000 ,      0.4668)
  (    6.0000 ,      0.3968)
  (    9.0000 ,      0.3155)
  (   12.0000 ,      0.2343)
  (   15.0000 ,      0.1562)
  (   18.0000 ,      0.1130)
  (   21.0000 ,      0.0887)
  (   24.0000 ,      0.0784)
  (   27.0000 ,      0.0641)
  (   30.0000 ,      0.0627)
};
\addplot[mark=diamond] coordinates{
  (    0.0000 ,      0.5434)
  (    3.0000 ,      0.4698)
  (    6.0000 ,      0.3838)
  (    9.0000 ,      0.2955)
  (   12.0000 ,      0.2296)
  (   15.0000 ,      0.1714)
  (   18.0000 ,      0.1451)
  (   21.0000 ,      0.1334)
  (   24.0000 ,      0.1323)
  (   27.0000 ,      0.1175)
  (   30.0000 ,      0.1168)
};
\addplot[mark=square] coordinates{
  (    0.0000 ,      0.6348)
  (    3.0000 ,      0.5965)
  (    6.0000 ,      0.5537)
  (    9.0000 ,      0.5148)
  (   12.0000 ,      0.4880)
  (   15.0000 ,      0.4721)
  (   18.0000 ,      0.4599)
  (   21.0000 ,      0.4588)
  (   24.0000 ,      0.4609)
  (   27.0000 ,      0.4487)
  (   30.0000 ,      0.4505)
};
\legend{Proposed VBI, Proposed GAMP, BODAMP, Optimal SOAV, Original SOAV, Standard SBL}
\end{semilogyaxis}
\end{tikzpicture}
~
\begin{tikzpicture}[scale=1]
\begin{semilogyaxis}[
ylabel={SER},grid=both, title={(b)},
legend style={at={(0.195,0.52),font=\footnotesize},
anchor=north,legend columns=1}, xmin=0,xmax=30,ymin=0.0004,width=10cm,height=5.88cm]
\addplot[mark=asterisk,red]  coordinates{
  (    0.0000 ,     0.7271)
  (    3.0000 ,     0.6733)
  (    6.0000 ,     0.6370)
  (    9.0000 ,     0.5869)
  (   12.0000 ,     0.5085)
  (   15.0000 ,     0.3992)
  (   18.0000 ,     0.2458)
  (   21.0000 ,     0.0424)
  (   24.0000 ,     0.0020)
  (   27.0000 ,          0)
  (   30.0000 ,          0)
};
\addplot[mark=o,blue]  coordinates{
  (    0.0000 ,     0.7836)
  (    3.0000 ,     0.7931)
  (    6.0000 ,     0.7930)
  (    9.0000 ,     0.7924)
  (   12.0000 ,     0.7941)
  (   15.0000 ,     0.8016)
  (   18.0000 ,     0.7976)
  (   21.0000 ,     0.7952)
  (   24.0000 ,     0.7949)
  (   27.0000 ,     0.7937)
  (   30.0000 ,     0.7966)
};
\addplot[mark=triangle]  coordinates{
  (    0.0000 ,      0.8367)
  (    3.0000 ,      0.8409)
  (    6.0000 ,      0.8374)
  (    9.0000 ,      0.8363)
  (   12.0000 ,      0.8390)
  (   15.0000 ,      0.8348)
  (   18.0000 ,      0.8391)
  (   21.0000 ,      0.8420)
  (   24.0000 ,      0.8406)
  (   27.0000 ,      0.8356)
  (   30.0000 ,      0.8370)
};
\addplot[mark=+] coordinates{
  (    0.0000 ,       0.7307)
  (    3.0000 ,       0.6877)
  (    6.0000 ,       0.6342)
  (    9.0000 ,       0.5622)
  (   12.0000 ,       0.4788)
  (   15.0000 ,       0.3778)
  (   18.0000 ,       0.2880)
  (   21.0000 ,       0.2266)
  (   24.0000 ,       0.1930)
  (   27.0000 ,       0.1589)
  (   30.0000 ,       0.1615)
};
\addplot[mark=diamond] coordinates{
  (    0.0000 ,      0.7453)
  (    3.0000 ,      0.6931)
  (    6.0000 ,      0.6202)
  (    9.0000 ,      0.5432)
  (   12.0000 ,      0.4746)
  (   15.0000 ,      0.4230)
  (   18.0000 ,      0.3816)
  (   21.0000 ,      0.3644)
  (   24.0000 ,      0.3502)
  (   27.0000 ,      0.3396)
  (   30.0000 ,      0.3388)
};
\addplot[mark=square] coordinates{
  (    0.0000 ,     0.7672)
  (    3.0000 ,     0.7514)
  (    6.0000 ,     0.7187)
  (    9.0000 ,     0.7008)
  (   12.0000 ,     0.6810)
  (   15.0000 ,     0.6607)
  (   18.0000 ,     0.6450)
  (   21.0000 ,     0.6362)
  (   24.0000 ,     0.6394)
  (   27.0000 ,     0.6378)
  (   30.0000 ,     0.6440)
};
\end{semilogyaxis}
\end{tikzpicture}
~
\begin{tikzpicture}[scale=1]
\begin{semilogyaxis}[xlabel={SNR [dB]},
ylabel={SER},grid=both, title={(c)},
legend style={at={(0.25,0.52),font=\footnotesize},
anchor=north,legend columns=1}, xmin=0,xmax=30,ymin=0.0004,width=10cm,height=5.88cm]
\addplot[mark=asterisk,red]  coordinates{
  (    0.0000 ,      0.8614)
  (    3.0000 ,      0.8300)
  (    6.0000 ,      0.7694)
  (    9.0000 ,      0.7454)
  (   12.0000 ,      0.6972)
  (   15.0000 ,      0.6044)
  (   18.0000 ,      0.4725)
  (   21.0000 ,      0.3096)
  (   24.0000 ,      0.0652)
  (   27.0000 ,      0.0004)
  (   30.0000 ,           0)
};
\addplot[mark=o,blue]  coordinates{
  (    0.0000 ,      0.8811)
  (    3.0000 ,      0.8855)
  (    6.0000 ,      0.8833)
  (    9.0000 ,      0.8866)
  (   12.0000 ,      0.8896)
  (   15.0000 ,      0.8890)
  (   18.0000 ,      0.8883)
  (   21.0000 ,      0.8862)
  (   24.0000 ,      0.8869)
  (   27.0000 ,      0.8924)
  (   30.0000 ,      0.8856)
};
\addplot[mark=triangle]  coordinates{
  (    0.0000 ,      0.9157)
  (    3.0000 ,      0.9135)
  (    6.0000 ,      0.9176)
  (    9.0000 ,      0.9170)
  (   12.0000 ,      0.9160)
  (   15.0000 ,      0.9193)
  (   18.0000 ,      0.9205)
  (   21.0000 ,      0.9169)
  (   24.0000 ,      0.9220)
  (   27.0000 ,      0.9144)
  (   30.0000 ,      0.9190)
};
\addplot[mark=+] coordinates{
  (    0.0000 ,       0.8537)
  (    3.0000 ,       0.8261)
  (    6.0000 ,       0.7931)
  (    9.0000 ,       0.7412)
  (   12.0000 ,       0.6724)
  (   15.0000 ,       0.5878)
  (   18.0000 ,       0.4959)
  (   21.0000 ,       0.4044)
  (   24.0000 ,       0.3223)
  (   27.0000 ,       0.2721)
  (   30.0000 ,       0.2524)
};
\addplot[mark=diamond] coordinates{
  (    0.0000 ,      0.8659)
  (    3.0000 ,      0.8345)
  (    6.0000 ,      0.7827)
  (    9.0000 ,      0.7285)
  (   12.0000 ,      0.6731)
  (   15.0000 ,      0.6185)
  (   18.0000 ,      0.5898)
  (   21.0000 ,      0.5785)
  (   24.0000 ,      0.5612)
  (   27.0000 ,      0.5551)
  (   30.0000 ,      0.5531)
};
\addplot[mark=square] coordinates{
  (    0.0000 ,      0.8603)
  (    3.0000 ,      0.8434)
  (    6.0000 ,      0.8276)
  (    9.0000 ,      0.8114)
  (   12.0000 ,      0.7967)
  (   15.0000 ,      0.7889)
  (   18.0000 ,      0.7751)
  (   21.0000 ,      0.7752)
  (   24.0000 ,      0.7691)
  (   27.0000 ,      0.7695)
  (   30.0000 ,      0.7716)
};
\end{semilogyaxis}
\end{tikzpicture}
\caption{SER  versus SNR with a correlated measurement matrix and $N=100$. a) $\Delta=0.7$ and $L=4$; b)  $\Delta=0.8$ and $L=8$; c) $\Delta=0.9$ and $L=16$.
 }\label{fig-SNR-C}
\end{figure}
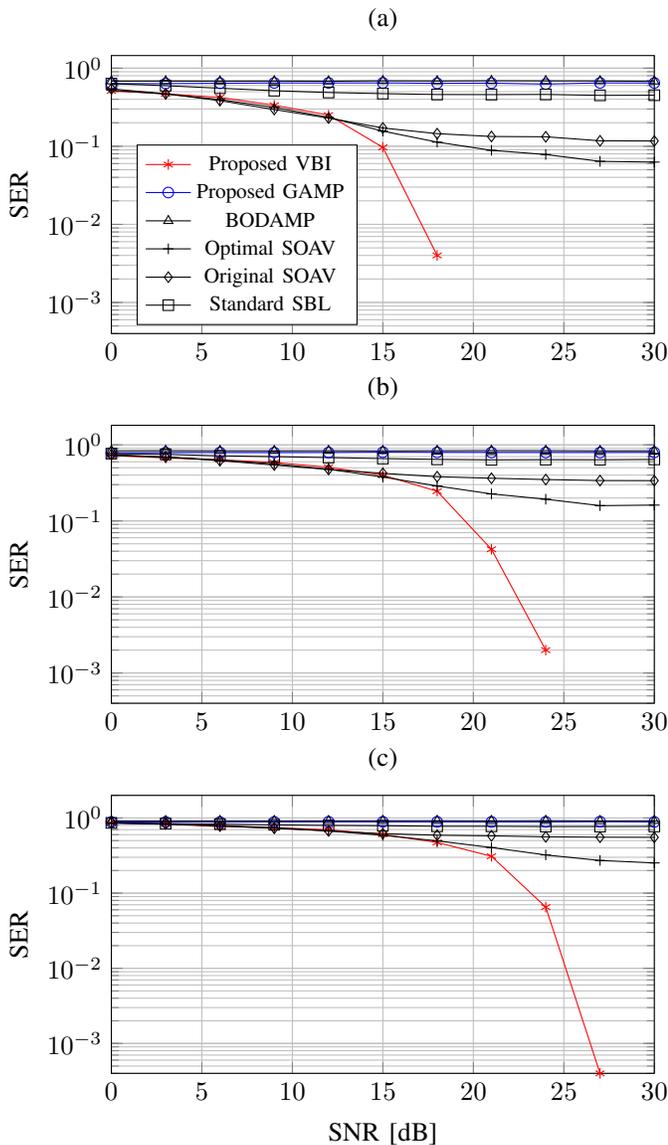

In Figs.~\ref{fig-SNR-G} and \ref{fig-SNR-C}, Monte Carlo trials are carried out to investigate the impact of the signal-to-noise ratio (SNR) on the
symbol error rate (SER) performance, where the SER is defined as
\begin{align}
\frac{1}{M_c N}\sum_{m=1}^{M_c} {\left\| \x^{\mathrm{est},m} - \x^{\mathrm{true}} \right\|_0 },
\end{align}
where 
$\x^{\mathrm{est},m}$ is the estimate of $\x^{\mathrm{true}}$ at the $m$-th Monte Carlo trial and $M_c=200$ is the number of trials.
The maximum number of iterations for each strategy is fixed to 100.
Fig.~\ref{fig-SNR-G} shows the SER performance of the discrete signal reconstruction based on different
strategies with an i.i.d. Gaussian measurement matrix versus SNR.
Fig.~\ref{fig-SNR-C} shows the SER performance of the discrete signal reconstruction with a correlated measurement matrix versus SNR.
It is seen that (i) the GAMP-based method again gives the best performance with an i.i.d. Gaussian measurement matrix, but fails to work with a non i.i.d. Gaussian measurement matrix; (ii) the  VBI-based method always retains a reasonable  SER performance with either an i.i.d. Gaussian measurement matrix or a correlated measurement matrix; and (iii) BODAMP can achieve a good SER performance with an i.i.d. Gaussian measurement matrix but it also fails to work with a correlated measurement matrix; and (iv) the optimal SOAV method outperforms the original SOAV method, but it is inferior to the VBI-based method.

\subsection{SER Performance Versus $L$}

\begin{figure}
\center
\begin{tikzpicture}[scale=1]
\begin{semilogyaxis}[
ylabel={SER},grid=both, title={(a)},
legend style={at={(0.805,0.66),font=\footnotesize},
anchor=north,legend columns=1}, xmin=2,xmax=16,width=10cm,height=5.88cm]
\addplot[mark=asterisk,red]  coordinates{
  (        2,            0   )   
  (        4,            0   )   
  (        6,       0.0048   )   
  (        8,       0.0776   )   
  (       10,       0.2250   )   
  (       12,       0.3129   )   
  (       14,       0.3887   )   
  (       16,       0.4215   )   
};
\addplot[mark=o,blue]  coordinates{
  (        2,     0.3936)   
  (        4,     0.6121)   
  (        6,     0.7093)   
  (        8,     0.7648)   
  (       10,     0.8012)   
  (       12,     0.8330)   
  (       14,     0.8474)   
  (       16,     0.8695)   
};
\addplot[mark=triangle]  coordinates{
  (        2,     0.4108)   
  (        4,     0.6843)   
  (        6,     0.7834)   
  (        8,     0.8375)   
  (       10,     0.8682)   
  (       12,     0.8912)   
  (       14,     0.9076)   
  (       16,     0.9179)   
};
\addplot[mark=+] coordinates{
  (        2,     0.0012)   
  (        4,     0.0254)   
  (        6,     0.1443)   
  (        8,     0.2384)   
  (       10,     0.3481)   
  (       12,     0.4248)   
  (       14,     0.5091)   
  (       16,     0.5516)   
};
\addplot[mark=diamond] coordinates{
  (        2,     0.0100)   
  (        4,     0.0767)   
  (        6,     0.2382)   
  (        8,     0.3704)   
  (       10,     0.4524)   
  (       12,     0.5319)   
  (       14,     0.5851)   
  (       16,     0.6373)   
};
\addplot[mark=square] coordinates{
  (        2,     0.1507)   
  (        4,     0.3884)   
  (        6,     0.5478)   
  (        8,     0.6436)   
  (       10,     0.7121)   
  (       12,     0.7540)   
  (       14,     0.7906)   
  (       16,     0.8156)   
};
\legend{Proposed VBI, Proposed GAMP, BODAMP, Optimal SOAV, Original SOAV, Standard SBL, }
\end{semilogyaxis}
\end{tikzpicture}
~
\begin{tikzpicture}[scale=1]
\begin{semilogyaxis}[xlabel={$L$},
ylabel={SER},grid=both, title={(b)},
legend style={at={(0.22,0.09),font=\footnotesize},
anchor=north,legend columns=1}, xmin=2,xmax=16,width=10cm,height=5.88cm]
\addplot[mark=asterisk,red]  coordinates{
  (        2,           0)  
  (        4,           0)  
  (        6,      0.0003)  
  (        8,      0.0178)  
  (       10,      0.0887)  
  (       12,      0.2130)  
  (       14,      0.3028)  
  (       16,      0.3462)  
};
\addplot[mark=o,blue]  coordinates{
  (        2,           0)   
  (        4,           0)   
  (        6,           0)   
  (        8,           0)   
  (       10,      0.0001)   
  (       12,      0.0016)   
  (       14,      0.0110)   
  (       16,      0.0538)   
};
\addplot[mark=triangle]  coordinates{
  (        2,      0.0006  )  
  (        4,      0.0112  )  
  (        6,      0.0803  )  
  (        8,      0.1486  )  
  (       10,      0.2648  )  
  (       12,      0.3319  )  
  (       14,      0.4167  )  
  (       16,      0.4606  )  
};
\addplot[mark=+] coordinates{
  (        2,        0.0008)  
  (        4,        0.0143)  
  (        6,        0.1023)  
  (        8,        0.1966)  
  (       10,        0.2907)  
  (       12,        0.3750)  
  (       14,        0.4632)  
  (       16,        0.5096)  
};
\addplot[mark=diamond] coordinates{
  (        2,       0.0062 )  
  (        4,       0.0632 )  
  (        6,       0.2058 )  
  (        8,       0.3400 )  
  (       10,       0.4220 )  
  (       12,       0.5044 )  
  (       14,       0.5649 )  
  (       16,       0.6175 )  
};
\addplot[mark=square] coordinates{
  (        2,        0.1502)   
  (        4,        0.3865)   
  (        6,        0.5312)   
  (        8,        0.6394)   
  (       10,        0.7038)   
  (       12,        0.7491)   
  (       14,        0.7813)   
  (       16,        0.8129)   
};
\end{semilogyaxis}
\end{tikzpicture}
\caption{SER versus the number of elements in the finite alphabet $\mathcal{F}$, where  $N=100$, $\Delta=0.8$ and  SNR $=20$ dB. a) correlated measurement matrix; b) i.i.d. Gaussian measurement matrix.
 }\label{fig-L}
\end{figure}
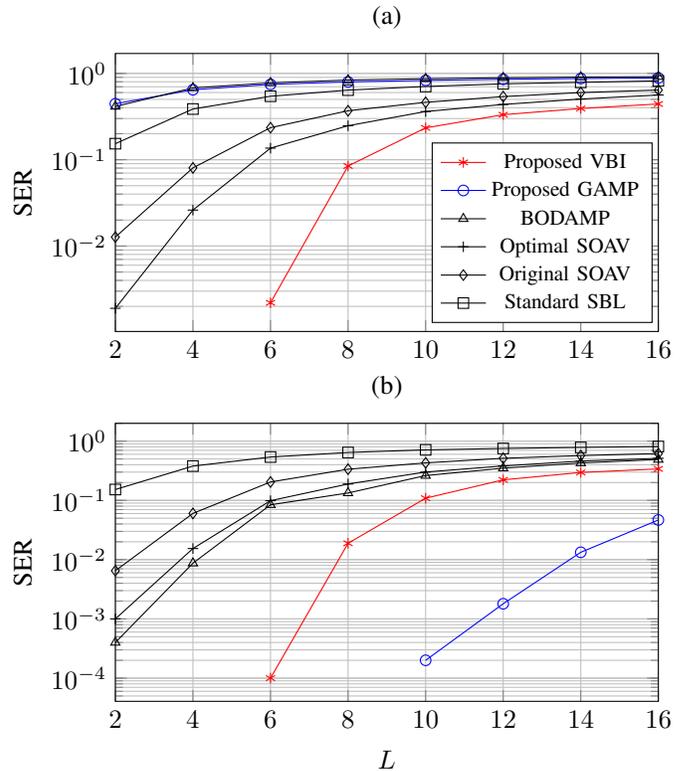

In Fig.~\ref{fig-L}, we study the impact of the size of the finite alphabet $\mathcal{F}$ on the SER performance.
Assume that $N=100$, $\Delta=0.8$ and SNR is set to 20 dB.
Fig.~\ref{fig-L} shows the SER  of the discrete signal reconstruction
versus the number of elements in the finite alphabet $\mathcal{F}$.
It is observed that(i) the SERs of all the methods increase as $L$ increases, because the distance between the two nearby elements in the finite alphabet $\mathcal{F}$ becomes small which will definitely cause a high SER;
(ii) the simulation results reconfirm that the GAMP-based scheme works perfectly with an i.i.d. Gaussian measurement matrix, and the type of measurement matrix does not affect the performance of the VBI-based method; and (iii) the VBI-based approach always outperforms the state-of-the-art methods.

\subsection{Success Rate Versus $\Delta$}

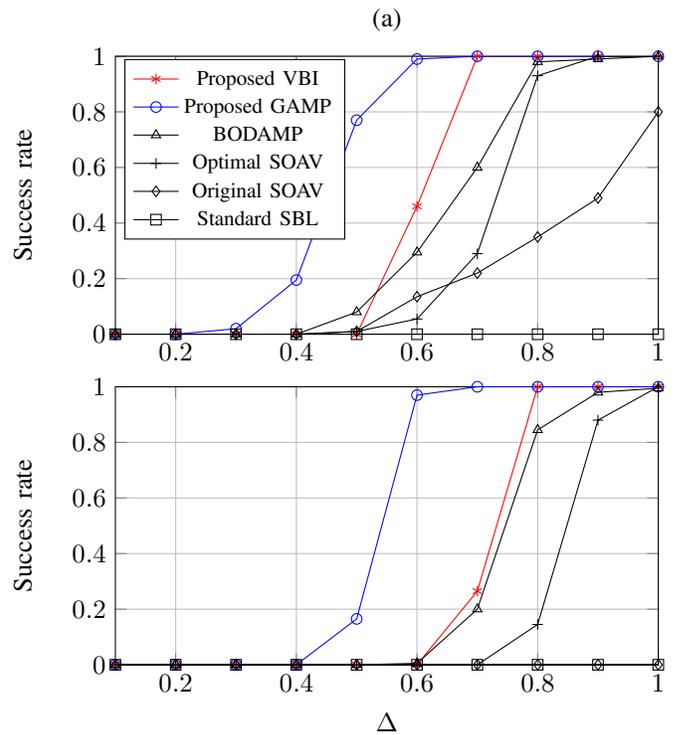
\begin{figure}
\center
\begin{tikzpicture}[scale=1]
\begin{axis}[
ylabel={Success rate},grid=both, title={(a)},
legend style={at={(0.195,0.99),font=\footnotesize},
anchor=north,legend columns=1}, xmin=0.1,xmax=1,ymin=0,ymax=1,width=10cm,height=5.88cm]
\addplot[mark=asterisk,red]  coordinates{
  (     0.1000 ,           0)
  (     0.2000 ,           0)
  (     0.3000 ,           0)
  (     0.4000 ,           0)
  (     0.5000 ,           0)
  (     0.6000 ,      0.4600)
  (     0.7000 ,      1.0000)
  (     0.8000 ,      1.0000)
  (     0.9000 ,      1.0000)
  (     1.0000 ,      1.0000)
};
\addplot[mark=o,blue]  coordinates{
  (     0.1000 ,           0)
  (     0.2000 ,           0)
  (     0.3000 ,      0.0200)
  (     0.4000 ,      0.1950)
  (     0.5000 ,      0.7700)
  (     0.6000 ,      0.9900)
  (     0.7000 ,      1.0000)
  (     0.8000 ,      1.0000)
  (     0.9000 ,      1.0000)
  (     1.0000 ,      1.0000)
};
\addplot[mark=triangle]  coordinates{
  (     0.1000 ,            0)
  (     0.2000 ,            0)
  (     0.3000 ,            0)
  (     0.4000 ,            0)
  (     0.5000 ,       0.0800)
  (     0.6000 ,       0.2950)
  (     0.7000 ,       0.6000)
  (     0.8000 ,       0.9800)
  (     0.9000 ,       0.9900)
  (     1.0000 ,       1.0000)
};
\addplot[mark=+] coordinates{
  (     0.1000 ,           0)
  (     0.2000 ,           0)
  (     0.3000 ,           0)
  (     0.4000 ,           0)
  (     0.5000 ,      0.0100)
  (     0.6000 ,      0.0550)
  (     0.7000 ,      0.2900)
  (     0.8000 ,      0.9300)
  (     0.9000 ,      1.0000)
  (     1.0000 ,      1.0000)
};
\addplot[mark=diamond] coordinates{
  (     0.1000 ,           0)
  (     0.2000 ,           0)
  (     0.3000 ,           0)
  (     0.4000 ,           0)
  (     0.5000 ,      0.0100)
  (     0.6000 ,      0.1350)
  (     0.7000 ,      0.2200)
  (     0.8000 ,      0.3500)
  (     0.9000 ,      0.4900)
  (     1.0000 ,      0.8000)
};
\addplot[mark=square] coordinates{
  (     0.1000 ,          0)
  (     0.2000 ,          0)
  (     0.3000 ,          0)
  (     0.4000 ,          0)
  (     0.5000 ,          0)
  (     0.6000 ,          0)
  (     0.7000 ,          0)
  (     0.8000 ,          0)
  (     0.9000 ,          0)
  (     1.0000 ,          0)
};
\legend{Proposed VBI, Proposed GAMP, BODAMP, Optimal SOAV, Original SOAV, Standard SBL}
\end{axis}
\end{tikzpicture}
~
\begin{tikzpicture}[scale=1]
\begin{axis}[xlabel={$\Delta$},
ylabel={Success rate},grid=both, 
legend style={at={(0.22,0.99),font=\footnotesize},
anchor=north,legend columns=1}, xmin=0.1,xmax=1,ymin=0,ymax=1,width=10cm,height=5.88cm]
\addplot[mark=asterisk,red]  coordinates{
  (     0.1000 ,           0)
  (     0.2000 ,           0)
  (     0.3000 ,           0)
  (     0.4000 ,           0)
  (     0.5000 ,           0)
  (     0.6000 ,           0)
  (     0.7000 ,      0.2650)
  (     0.8000 ,      1.0000)
  (     0.9000 ,      1.0000)
  (     1.0000 ,      1.0000)
};
\addplot[mark=o,blue]  coordinates{
  (     0.1000 ,           0)
  (     0.2000 ,           0)
  (     0.3000 ,           0)
  (     0.4000 ,           0)
  (     0.5000 ,      0.1650)
  (     0.6000 ,      0.9700)
  (     0.7000 ,      1.0000)
  (     0.8000 ,      1.0000)
  (     0.9000 ,      1.0000)
  (     1.0000 ,      1.0000)
};
\addplot[mark=triangle]  coordinates{
  (     0.1000 ,           0)
  (     0.2000 ,           0)
  (     0.3000 ,           0)
  (     0.4000 ,           0)
  (     0.5000 ,           0)
  (     0.6000 ,      0.0050)
  (     0.7000 ,      0.2000)
  (     0.8000 ,      0.8450)
  (     0.9000 ,      0.9800)
  (     1.0000 ,      0.9950)
};
\addplot[mark=+] coordinates{
  (     0.1000 ,           0)
  (     0.2000 ,           0)
  (     0.3000 ,           0)
  (     0.4000 ,           0)
  (     0.5000 ,           0)
  (     0.6000 ,           0)
  (     0.7000 ,           0)
  (     0.8000 ,      0.1450)
  (     0.9000 ,      0.8800)
  (     1.0000 ,      1.0000)
};
\addplot[mark=diamond] coordinates{
  (     0.1000 ,        0)
  (     0.2000 ,        0)
  (     0.3000 ,        0)
  (     0.4000 ,        0)
  (     0.5000 ,        0)
  (     0.6000 ,        0)
  (     0.7000 ,        0)
  (     0.8000 ,        0)
  (     0.9000 ,        0)
  (     1.0000 ,        0)
};
\addplot[mark=square] coordinates{
  (     0.1000 ,        0)
  (     0.2000 ,        0)
  (     0.3000 ,        0)
  (     0.4000 ,        0)
  (     0.5000 ,        0)
  (     0.6000 ,        0)
  (     0.7000 ,        0)
  (     0.8000 ,        0)
  (     0.9000 ,        0)
  (     1.0000 ,        0)
};
\end{axis}
\end{tikzpicture}
\caption{Success rate versus $\Delta$ with an i.i.d. Gaussian measurement matrix and $N=100$ int the noise-free case. a) $L=4$; b) $L=8$.
 }\label{fig-Delta-G}
\end{figure}

In Figs.~\ref{fig-Delta-G} and \ref{fig-Delta-C}, Monte Carlo trials are carried out to investigate the rate of the success recovery, which is defined as
\begin{align}
   \frac{1}{M_c}\sum_{m=1}^{M_c} {[\x^{\mathrm{est},m} = \x^{\mathrm{true}}]}.
\end{align}
Fig.~\ref{fig-Delta-G} shows the success rate of the discrete signal reconstruction with an i.i.d. Gaussian measurement matrix versus $\Delta$.
Fig.~\ref{fig-Delta-C} shows the success rate of the discrete signal reconstruction with a correlated measurement matrix versus $\Delta$. Note that we consider the noise-free case in both figures. 
It is seen that (i) the success rates of all the methods increase as $\Delta$ increases; (ii) the  GAMP-based method still provides the best performance with an i.i.d. Gaussian measurement matrix; and (iii) compared with the GAMP-based method, there is a little performance loss for the VBI-based scheme, but it retains a good success rate with a correlated measurement matrix and always outperforms the state-of-the-art methods.

\subsection{Runtime Versus $N$}
Finally, we carry out the computational complexity comparison versus the dimension of discrete signal $N$, where $\Delta=0.8$, $L=8$, and SNR is set to 20 dB. Fig.~\ref{fig-NN}-a shows the average runtime over 200 Monte Carlo trials  with an i.i.d. Gaussian measurement matrix, and Fig.~\ref{fig-NN}-b shows the corresponding SER performance for reference. The element-independent factorization method (named SAVE) is additionally included in each sub-figure. Note that the difference between SAVE and the proposed VBI-based method is in the adopted approximation factorization for $\x$ only (as discussed in Section 3.3.2).
It is observed that (i) the runtime of all the methods increases with $N$; (ii) the VBI-based method and the standard SBL method are much slower than other methods, but the VBI-based method can achieve the best SER performance; (iii) SAVE can provide a fast solution, but it suffers from a performance loss; (iv) the GAMP-based method can significantly reduce the computational complexity, and has very similar runtime as the SOAV-type methods; and (v) the GAMP-based method  achieves almost the perfect SER performance with an i.i.d. Gaussian measurement matrix, whose curve is out of the range of Fig.~\ref{fig-NN}-b.

\begin{figure}
\center
\begin{tikzpicture}[scale=1]
\begin{axis}[
ylabel={Success rate},grid=both, title={(a)},
legend style={at={(0.195,0.99),font=\footnotesize},
anchor=north,legend columns=1}, xmin=0.1,xmax=1,ymin=0,ymax=1,width=10cm,height=5.88cm]
\addplot[mark=asterisk,red]  coordinates{
  (   0.1000,           0)
  (   0.2000,           0)
  (   0.3000,           0)
  (   0.4000,           0)
  (   0.5000,      0.4950)
  (   0.6000,      1.0000)
  (   0.7000,      1.0000)
  (   0.8000,      1.0000)
  (   0.9000,      1.0000)
  (   1.0000,      1.0000)
};
\addplot[mark=o,blue]  coordinates{
  (    0.1000,           0)
  (    0.2000,      0.0050)
  (    0.3000,      0.0050)
  (    0.4000,      0.0050)
  (    0.5000,           0)
  (    0.6000,           0)
  (    0.7000,      0.0050)
  (    0.8000,      0.0050)
  (    0.9000,           0)
  (    1.0000,           0)
};
\addplot[mark=triangle]  coordinates{
  (   0.1000,        0)
  (   0.2000,        0)
  (   0.3000,        0)
  (   0.4000,        0)
  (   0.5000,        0)
  (   0.6000,        0)
  (   0.7000,        0)
  (   0.8000,        0)
  (   0.9000,        0)
  (   1.0000,        0)
};
\addplot[mark=+] coordinates{
  (    0.1000,            0)
  (    0.2000,            0)
  (    0.3000,            0)
  (    0.4000,            0)
  (    0.5000,            0)
  (    0.6000,       0.0100)
  (    0.7000,       0.2600)
  (    0.8000,       0.8700)
  (    0.9000,       1.0000)
  (    1.0000,       1.0000)
};
\addplot[mark=diamond] coordinates{
  (    0.1000,            0)
  (    0.2000,            0)
  (    0.3000,            0)
  (    0.4000,            0)
  (    0.5000,            0)
  (    0.6000,       0.0050)
  (    0.7000,       0.0250)
  (    0.8000,       0.2000)
  (    0.9000,       0.6100)
  (    1.0000,       0.8650)
};
\addplot[mark=square] coordinates{
  (    0.1000,         0)
  (    0.2000,         0)
  (    0.3000,         0)
  (    0.4000,         0)
  (    0.5000,         0)
  (    0.6000,         0)
  (    0.7000,         0)
  (    0.8000,         0)
  (    0.9000,         0)
  (    1.0000,         0)
};
\legend{Proposed VBI, Proposed GAMP, BODAMP, Optimal SOAV, Original SOAV, Standard SBL}
\end{axis}
\end{tikzpicture}
~
\begin{tikzpicture}[scale=1]
\begin{axis}[xlabel={$\Delta$},
ylabel={Success rate},grid=both, title={(b)},
legend style={at={(0.195,0.99),font=\footnotesize},
anchor=north,legend columns=1}, xmin=0.1,xmax=1,ymin=0,ymax=1,width=10cm,height=5.88cm]
\addplot[mark=asterisk,red]  coordinates{
  (   0.1000,           0)
  (   0.2000,           0)
  (   0.3000,           0)
  (   0.4000,           0)
  (   0.5000,           0)
  (   0.6000,           0)
  (   0.7000,      0.6900)
  (   0.8000,      1.0000)
  (   0.9000,      1.0000)
  (   1.0000,      1.0000)
};
\addplot[mark=o,blue]  coordinates{
  (    0.1000,       0)
  (    0.2000,       0)
  (    0.3000,       0)
  (    0.4000,       0)
  (    0.5000,       0)
  (    0.6000,       0)
  (    0.7000,       0)
  (    0.8000,       0)
  (    0.9000,       0)
  (    1.0000,       0)
};
\addplot[mark=triangle]  coordinates{
  (    0.1000,        0)
  (    0.2000,        0)
  (    0.3000,        0)
  (    0.4000,        0)
  (    0.5000,        0)
  (    0.6000,        0)
  (    0.7000,        0)
  (    0.8000,        0)
  (    0.9000,        0)
  (    1.0000,        0)
};
\addplot[mark=+] coordinates{
  (    0.1000,           0)
  (    0.2000,           0)
  (    0.3000,           0)
  (    0.4000,           0)
  (    0.5000,           0)
  (    0.6000,           0)
  (    0.7000,           0)
  (    0.8000,      0.0750)
  (    0.9000,      0.7450)
  (    1.0000,      1.0000)
};
\addplot[mark=diamond] coordinates{
  (   0.1000,            0)
  (   0.2000,            0)
  (   0.3000,            0)
  (   0.4000,            0)
  (   0.5000,            0)
  (   0.6000,            0)
  (   0.7000,            0)
  (   0.8000,            0)
  (   0.9000,            0)
  (   1.0000,       0.0050)
};
\addplot[mark=square] coordinates{
  (    0.1000,        0)
  (    0.2000,        0)
  (    0.3000,        0)
  (    0.4000,        0)
  (    0.5000,        0)
  (    0.6000,        0)
  (    0.7000,        0)
  (    0.8000,        0)
  (    0.9000,        0)
  (    1.0000,        0)
};
\end{axis}
\end{tikzpicture}
\caption{Success rate versus $\Delta$ with a correlated measurement matrix and $N=100$ in the noise-free case. a) $L=3$; b)
$L=6$.
 }\label{fig-Delta-C}
\end{figure}
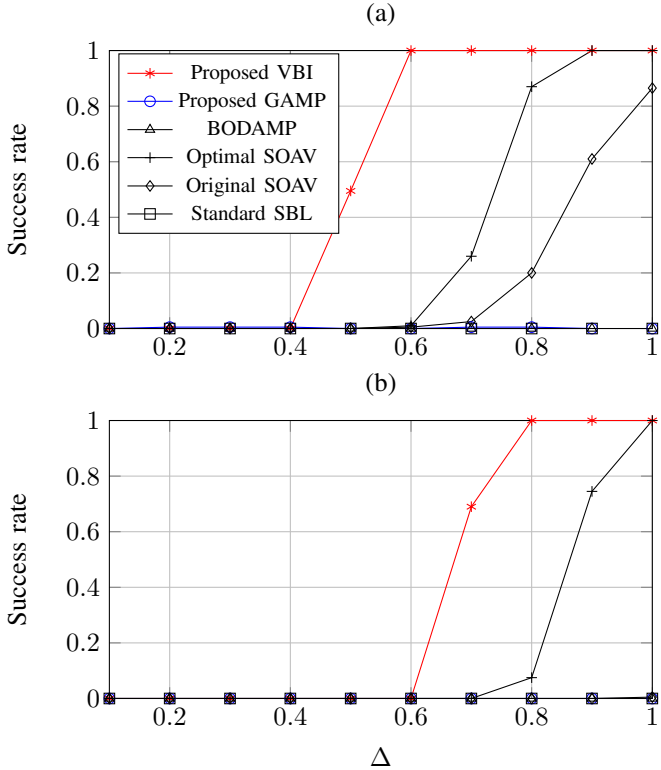

\begin{figure}  
\center
\begin{tikzpicture}[scale=1]
\begin{semilogyaxis}[
ylabel={Time [seconds]},grid=both, title={(a)},
legend style={at={(0.19,1.15),font=\footnotesize},
anchor=north,legend columns=1}, xmin=50,xmax=400,width=10cm,height=5.88cm]
\addplot[mark=asterisk,red]  coordinates{
  (    50,       0.0187)   
  (   100,       0.0510)   
  (   150,       0.1210)   
  (   200,       0.2414)   
  (   250,       0.4155)   
  (   300,       0.8002)   
  (   350,       1.2926)   
  (   400,       1.7306)   
};
\addplot[mark=o,blue]  coordinates{
  (    50,       0.0013)   
  (   100,       0.0021)   
  (   150,       0.0043)   
  (   200,       0.0057)   
  (   250,       0.0070)   
  (   300,       0.0088)   
  (   350,       0.0109)   
  (   400,       0.0162)   
};
\addplot[mark=triangle]  coordinates{
  (    50,       0.0034)   
  (   100,       0.0070)   
  (   150,       0.0102)   
  (   200,       0.0126)   
  (   250,       0.0150)   
  (   300,       0.0177)   
  (   350,       0.0219)   
  (   400,       0.0232)   
};
\addplot[mark=+] coordinates{
  (    50,        0.0016)   
  (   100,        0.0031)   
  (   150,        0.0038)   
  (   200,        0.0048)   
  (   250,        0.0058)   
  (   300,        0.0076)   
  (   350,        0.0103)   
  (   400,        0.0115)   
};
\addplot[mark=diamond] coordinates{
  (    50,       0.0016)   
  (   100,       0.0032)   
  (   150,       0.0038)   
  (   200,       0.0048)   
  (   250,       0.0059)   
  (   300,       0.0079)   
  (   350,       0.0104)   
  (   400,       0.0116)   
};
\addplot[mark=square] coordinates{
  (    50,      0.0116)   
  (   100,      0.0393)   
  (   150,      0.1036)   
  (   200,      0.2078)   
  (   250,      0.3751)   
  (   300,      0.7348)   
  (   350,      1.2135)   
  (   400,      1.5692)   
};
\addplot[mark=triangle*,violet] coordinates{
  (    50,       0.0166)   
  (   100,       0.0351)   
  (   150,       0.0664)   
  (   200,       0.1083)   
  (   250,       0.1483)   
  (   300,       0.2280)   
  (   350,       0.4997)   
  (   400,       0.5986)   
};
\legend{Proposed VBI, Proposed GAMP, BODAMP, Optimal SOAV, Original SOAV, Standard SBL, Proposed SAVE}
\end{semilogyaxis}
\end{tikzpicture}
~
\begin{tikzpicture}[scale=1]
\begin{semilogyaxis}[xlabel={$N$},
ylabel={SER},grid=both, title={(b)},
legend style={at={(0.22,0.09),font=\footnotesize},
anchor=north,legend columns=1}, xmin=50,xmax=400,width=10cm,height=5.88cm]
\addplot[mark=asterisk,red]  coordinates{
  (    50  ,      0.0524)  
  (   100  ,      0.0171)  
  (   150  ,      0.0140)  
  (   200  ,      0.0110)  
  (   250  ,      0.0077)  
  (   300  ,      0.0067)  
  (   350  ,      0.0047)  
  (   400  ,      0.0038)  
};
\addplot[mark=o,blue]  coordinates{
  (      50,         0)   
  (     100,         0)   
  (     150,         0)   
  (     200,         0)   
  (     250,         0)   
  (     300,         0)   
  (     350,         0)   
  (     400,         0)   
};
\addplot[mark=triangle]  coordinates{
  (    50,        0.1735)  
  (   100,        0.1321)  
  (   150,        0.1306)  
  (   200,        0.1185)  
  (   250,        0.1186)  
  (   300,        0.1229)  
  (   350,        0.1092)  
  (   400,        0.1157)  
};
\addplot[mark=+] coordinates{
  (    50,        0.2030)  
  (   100,        0.1964)  
  (   150,        0.1889)  
  (   200,        0.1942)  
  (   250,        0.1981)  
  (   300,        0.1987)  
  (   350,        0.1925)  
  (   400,        0.1905)  
};
\addplot[mark=diamond] coordinates{
  (    50,       0.3377)  
  (   100,       0.3333)  
  (   150,       0.3300)  
  (   200,       0.3410)  
  (   250,       0.3390)  
  (   300,       0.3367)  
  (   350,       0.3416)  
  (   400,       0.3373)  
};
\addplot[mark=square] coordinates{
  (    50,       0.6396)   
  (   100,       0.6261)   
  (   150,       0.6396)   
  (   200,       0.6354)   
  (   250,       0.6428)   
  (   300,       0.6380)   
  (   350,       0.6410)   
  (   400,       0.6349)   
};
\addplot[mark=triangle*,violet] coordinates{
  (    50,       0.0969)   
  (   100,       0.0555)   
  (   150,       0.0515)   
  (   200,       0.0447)   
  (   250,       0.0310)   
  (   300,       0.0311)   
  (   350,       0.0240)   
  (   400,       0.0194)   
};
\end{semilogyaxis}
\end{tikzpicture}
\caption{Runtime and SER versus the dimension of discrete signal with an i.i.d. Gaussian measurement matrix, $\Delta=0.8$, $L=8$, and  SNR $=20$ dB.
 }\label{fig-NN}
\end{figure}

\section{Conclusion}

The discrete signal reconstruction problem is tackled in this paper from the perspective of SBL.
Since  the ideal discretization prior (\ref{fa2}) is composed of several Dirac delta functions, it is usually
intractable to perform the Bayesian inference with (\ref{fa2}). To obtain a tractable Bayesian inference, we provide a novel discretization enforcing prior (\ref{eq-prix}) to exploit the knowledge of the discrete nature of the SOI.
Then, we  combine the new prior (\ref{eq-prix}) into the SBL framework and resort the VBI methodology to jointly characterize the finite-alphabet feature and reconstruct the unknown signal. Finally, we propose a fast GAMP-based method  to exploit the ideal discretization prior directly, as well as to reduce the computational burden significantly, in the presence of i.i.d. Gaussian measurement matrices.  Simulation results show that the VBI-based solution always outperforms the state-of-the-art SOAV optimization methods, and the GAMP-based scheme can further improve the discrete signal reconstruction performance if the measurement matrix is i.i.d. Gaussian.  However, for non i.i.d. Gaussian
measurement matrices, the GAMP-based method will fail to work; while the VBI-based method
with the new  prior (\ref{eq-prix}) does not require any assumption about the measurement matrix.

%
%
%
%
%



\appendix

\section{Proof of Lemma~1}

The following proof is similar to the one in \cite{dai2019non}. Let $C_\alpha\triangleq\{ b_\alpha \}$, $C_x=\{\bm\mu, \bm\Sigma \}$,  $C_\gamma\triangleq\{ b_n \}_{n=1}^N$ and $C_G \triangleq \{\phi_{n,l}\}_{n=1, l=1}^{N,L}$.
According to (\ref{eq-alg1}), (\ref{eq-alg2}), (\ref{eq-alg3}) and (\ref{eq-alg4}), each factor in $q(\bm\Omega)= q(\alpha) q(\x) q(\bm\gamma) q(\G)$ can be considered as a parameterized function, i.e.,
\begin{align}
q(\alpha)=& \Gamma(\alpha| C_\alpha ),\\
q(\x)=&  \mathcal{CN}(\x| C_x ),\\
q(\bm\gamma)=& \Gamma\left(\bm\gamma |C_\gamma  \right),
\end{align}
and $q(\G)$ is a discrete distribution parameterized by  $C_G$. Therefore, the functional optimization problem (\ref{eq-KLproblem}) can be formulated as a parameterized optimization problem
\begin{align}
\{C_\alpha^\star,  C_x^\star, C_\gamma^\star, C_G^\star   \} =  \min_{C_\alpha,  C_x, C_\gamma, C_G} D_{\mathrm{KL}}(C_\alpha,  C_x, C_\gamma, C_G) \label{eqparaproblem}
\end{align}
where $ D_{\mathrm{KL}}(C_\alpha,  C_x, C_\gamma, C_G)$ is the parameterized objective  function for $D_{\mathrm{KL}}(q(\bm\Omega )|| p(\bm\Omega| \y)  )$. Then, (\ref{eqM1})--(\ref{eqM4}) become:
\begin{align}
C_\alpha^{(i+1)}&= \arg \min_{C_\alpha}   D_{\mathrm{KL}}\left( C_\alpha,C_x^{(i)},C_\gamma^{(i)},C_G^{(i)} \right),\label{eqU1}\\
C_x^{(i+1)}&= \arg \min_{C_x}   D_{\mathrm{KL}}\left( C_\alpha^{(i+1)},C_x,C_\gamma^{(i)},C_G^{(i)} \right),\label{eqU2}\\
C_\gamma^{(i+1)}&= \arg \min_{C_\gamma}  D_{\mathrm{KL}}\left( C_\alpha^{(i+1)},C_x^{(i+1)},C_\gamma,C_G^{(i)} \right),\label{eqU3}\\
C_G^{(i+1)}&= \arg \min_{C_G}  D_{\mathrm{KL}}\left( C_\alpha^{(i+1)},C_x^{(i+1)},C_\gamma^{(i+1)},C_G \right).\label{eqU4}
\end{align}

Note that each subproblem has a unique solution, given in  (\ref{eq-alg1}), (\ref{eq-alg2}), (\ref{eq-alg3}) and (\ref{eq-alg4}). According to Theorem 2-b in \cite{razaviyayn2014successive}, the iterates generated by (\ref{eqU1})--(\ref{eqU4}) converge to a stationary point of the problem (\ref{eqparaproblem}) or, equivalently, (\ref{eq-KLproblem}).

\section*{References}

\bibliographystyle{elsarticle-num}
\bibliography{DiscreteSBL}

\end{document}